\RequirePackage{ifpdf}
\ifpdf 
\documentclass[pdftex]{sigma}
\else
\documentclass{sigma}
\fi

\numberwithin{equation}{section}
\numberwithin{theorem}{section}
\numberwithin{proposition}{section}
\numberwithin{lemma}{section}
\numberwithin{corollary}{section}
\numberwithin{definition}{section}
\numberwithin{example}{section}
\numberwithin{remark}{section}
\numberwithin{note}{section}

\usepackage{stmaryrd}
\usepackage{mathrsfs}
\usepackage{wasysym}
\def\frac#1#2{{#1\over#2}}

\def\M{\mathcal{M}}

\def\X{\mathcal{X}}
\def\U{\mathcal{U}}
\def\H{\mathcal{H}}
\def\A{\mathcal{A}}

\begin{document}

\allowdisplaybreaks

\renewcommand{\thefootnote}{$\star$}

\renewcommand{\PaperNumber}{086}

\FirstPageHeading

\ShortArticleName{$\kappa$-Minkowski Spacetimes and DSR Algebras: Fresh Look and Old Problems}

\ArticleName{$\boldsymbol{\kappa}$-Minkowski Spacetimes and DSR Algebras:\\ Fresh Look and Old Problems\footnote{This paper is a
contribution to the Special Issue ``Noncommutative Spaces and Fields''. The
full collection is available at
\href{http://www.emis.de/journals/SIGMA/noncommutative.html}{http://www.emis.de/journals/SIGMA/noncommutative.html}}}

\Author{Andrzej BOROWIEC and Anna PACHO{\L}}

\AuthorNameForHeading{A.~Borowiec and A.~Pacho{\l}}
\Address{Institute for Theoretical Physics, University of Wroclaw,\\
pl. Maxa Borna 9, 50-204 Wroc{\l}aw, Poland}
\Email{\href{mailto:borow@ift.uni.wroc.pl}{borow@ift.uni.wroc.pl}, \href{mailto:anna.pachol@ift.uni.wroc.pl}{anna.pachol@ift.uni.wroc.pl}}

\ArticleDates{Received March 30, 2010, in f\/inal form October 10, 2010;  Published online October 20, 2010}

\Abstract{Some classes of Deformed Special Relativity (DSR) theories are reconsidered within the Hopf algebraic formulation. For this purpose we shall explore a minimal framework of deformed Weyl--Heisenberg algebras provided by a smash product construction of DSR algebra. It is proved that this DSR algebra, which uniquely unif\/ies $\kappa$-Minkowski spacetime coordinates with Poincar\'{e} generators, can be obtained by nonlinear change of generators from undeformed one. Its various realizations in terms of the standard (undeformed) Weyl--Heisenberg algebra opens the way for quantum mechanical interpretation of DSR theories in terms of relativistic (St\"{u}ckelberg version) Quantum Mechanics. On this basis we review some recent results concerning twist realization of $\kappa$-Minkowski spacetime described as a quantum covariant algebra determining a deformation quantization of the corresponding linear Poisson structure.
Formal and conceptual issues concerning quantum $\kappa$-Poincar\'{e} and $\kappa$-Minkowski algebras as well as DSR theories are discussed. Particularly, the so-called ``$q$-analog'' version of DSR algebra is introduced. Is deformed special relativity quantization of doubly special relativity remains an open question. Finally, possible phy\-si\-cal applications of DSR algebra to description of some aspects of Planck scale physics are shortly recalled.}


\Classification{16T05; 17B37; 46L65; 53D55; 81R50; 81R60; 81T75; 83C65}


\renewcommand{\thefootnote}{\arabic{footnote}}
\setcounter{footnote}{0}

\section{Introduction}

$\kappa$-Minkowski spacetime \cite{Z,MR,Luk2}  is one of the examples of noncommutative spacetimes which has been studied extensively \cite{Z,MR,Luk2,Luk1,Kos,KosLuk,LukNow,Nowicki,GNbazy,Group21,Ballesteros,s1,Bu,Meljanac0702215,MSSKG,BP,BP2,BGMP,Dabrowski} for more than decade now and is interesting from physical and mathematical point of view. At f\/irst, noncommutative spacetimes~\cite{Dop,Oeckl,ABDMSW,AJSW,MSSW,JSSW,JMSSW,ADMSW,ADMW,Szabo2006,Chai1,Chai2,DJMTWW,FKGN,MS2,Dimitrijevic} in general may be useful in description of physics at the Planck scale since this energy scale might witness more general spacetime structure, i.e.\ noncommutative one. In this approach quantum uncertainty relations and discretization naturally arise~\cite{Dop}. $\kappa$-Minkowski spacetime is an especially interesting example because it is one of the possible frameworks for deformed special relativity theories (DSR), originally called doubly special relativity \cite{AC,BACKG,Smolin,Girelli,DSRkrytyka}. Together with this connection a physical interpretation for deformation parameter~$\kappa$,   as second invariant scale, appears naturally and allows us to interpret deformed dispersion relations as
valid at the ``$\kappa$-scale'' (as Planck scale or Quantum Gravity scale) when quantum gravity corrections become relevant.
Recent studies show that DSR theories might be experimentally falsif\/iable \cite{smolin1,Liberati,ACdsrmyth,DSRdebateI,DSRdebate}, which places $\kappa$-Minkowski spacetime on the frontier between strictly mathe\-ma\-tical theoretical example of noncommutative algebra and physical theory describing Planck scale nature.
 From the latter point of view noncommutative spacetimes are very interesting objects to work on and they are connected with quantum deformation techniques \cite{Drinfeld}. Quantum deformations which lead to noncommutative spacetimes are strictly connected with quantum groups \cite{Drinfeld,Jimbo,Fadeev,Woronowicz} generalizing symmetry groups.
In this way $\kappa$-Minkowski spacetime and $\kappa$-Poincar\'e algebra are related by the notion
of module algebra ($\equiv$ covariant quantum space \cite{Oeckl,DJMTWW,covQS})~--  algebraic generalization of covariant space.
To this aim one needs $\kappa$-Poincar\'{e} Hopf algebra. It has been originally discovered
in the so-called standard, inherited from anti-de
Sitter Lie algebra by contraction procedure, basis~\cite{Luk1}. Later on it has been reformulated by introducing easier bicrossproduct basis~\cite{MR}. Recently, the easiest basis determined by original classical Poincar\'{e} generators has being popularized to work with (see~\cite{BP2}, for earlier references see, e.g.~\cite{Kos,KosLuk,GNbazy}).
One can extend $\kappa$-Poincar\'{e} algebra by $\kappa$-Minkowski commutation relations using crossed (smash) product construction. Particularly this contains deformation of Weyl subalgebra, which is crossproduct of $\kappa$-Minkowski algebra with algebra of four momenta. One should notice that there have been other constructions of $\kappa$-Minkowski algebra ($\kappa$-Poincar\'{e} algebra) extension, by introducing $\kappa$-deformed phase space, e.g., in~\cite{LukNow, ncphasespace,AF} (see also~\cite{JKG}, the name ``DSR algebra'' was f\/irstly proposed here). However we would like to point out that Heisenberg double construction is not the only way to obtain DSR algebra. Particularly this leads to a deformation generated by momenta and coordinates of Weyl (Heisenberg) subalgebra. One of the advantages of using smash-product construction is that we leave an open geometrical interpretation of $\kappa$-Minkowski spacetime.
 Examples of $\kappa$-deformed phase space obtained by using crossproduct construction can be found also in~\cite{Kos,Nowicki}, however only for one specif\/ic realization related with the bicrossproduct \mbox{basis}.

\looseness=1
The present paper comprises  both research and review aspects. We review some results on va\-rious realizations of $\kappa$-Minkowski spacetime, however we also analyze its possible def\/initions and mathematical properties with more details. We also investigate new aspects of smash product algebras as pseudo-deformations. We describe dif\/ferent versions of quantum Minkowski spacetime algebra; twisted equipped with $h$-adic topology and covariant with respect to quantized general linear group (Section~\ref{section3}), $h$-adic twist independent and  covariant with respect to $\kappa$-Poincar\'{e} group (Section~\ref{section4.1}) and last but not least its the so-called $q$-analog version (Section~\ref{section4.2}). We believe that our approach might provide a better understanding and new perspective to the subject.

For the sake of completeness we start this paper with recalling few mathematical facts about Hopf module algebras, smash-product construction and Heisenberg realization with many illustrative examples.
Basic ideas of the twist deformation are also reminded with explicit form of Jordanian and Abelian twists leading to $\kappa$-Minkowski algebra which is afterwards extended via crossed product in both cases. Therefore we obtain deformed phase spaces as subalgebras. Moreover we notice that for any Drinfeld twist the twisted smash product algebra is isomorphic to undeformed one, hence we propose to call it a ``pseudo-deformation''. This statement is provided with the proof and Abelian and Jordanian cases are illustrative examples of it. Furthermore we drive the attention to the point that one can see $\kappa$-Minkowski spacetime as a quantization deformation in the line of Kontsevich~\cite{Kon}.
In the following (Section~\ref{section4}) we point out the main dif\/ferences between two versions of $\kappa$-Minkowski and $\kappa$-Poincar\'{e} algebras, $h$-adic and $q$-analog, and we construct DSR algebras for all of them with deformed phase spaces as subalgebras.  We show explicitly that DSR models depend upon  various Weyl algebra realizations  of $\kappa$-Minkowski spacetime one uses. The most interesting one from physical point of view seems to be version of $\kappa$-Minkowski spacetime with f\/ixed value of parameter $\kappa$ ($q$-analog). In this case $\kappa$-Minkowski algebra is an universal envelope of solvable Lie algebra without $h$-adic topology. This version allows us to connect the parameter $\kappa$ with some physical constant, like, e.g., quantum gravity scale or Planck mass and all the realizations might have physical interpretation. Also this version is used in Group Field Theories \cite{Oriti} which are connected with loop quantum gravity and spin foams approach. This part of (Section~\ref{section4}) the paper contains mostly new results.
Physical implications  and conclusions are summarized in the f\/inal section, where we explicitly show how dif\/ferent physical consequences, as dif\/ferent time delay predictions for photons or bounds on quantum gravity scale, appear.  Realizations of wide range of DSR algebra, in terms of undeformed Weyl algebra, lead to physically dif\/ferent models of DSR theory.
Our intention in this paper is to shed a light into technical aspects of $\kappa$-deformation, which were not properly treated in the physical literature, and provide correct def\/initions.

\section{Preliminaries: the Hopf-module algebra, smash product\\ and its Heisenberg representation}\label{section2}

Let us begin with the basic notion on a Hopf algebra considered as symmetry algebra (quantum group) of another algebra representing quantum space: the so-called  module algebra or  covariant quantum space. A~Hopf algebra is a complex\footnote{In this paper we shall work mainly with vector spaces over the f\/ield of complex numbers $\mathbb{C}$. All maps are
$\mathbb{C}$-linear maps.}, unital and associative algebra equipped with additional structures such as a comultiplication, a counit and an antipode. In some sense,
these structures and their axioms ref\/lect the multiplication, the unit element
and the inverse elements of a group and their corresponding properties \cite{Klimyk, Kassel}.

\begin{example}\label{example1}
Any Lie algebra $\mathfrak{g}$ provides an example of the (undeformed) Hopf algebra by taking its universal enveloping algebra $\U_{\mathfrak{g}}$
equipped with the primitive coproduct: $\Delta_0(u)=u\otimes 1+1\otimes u$, counit:  $\epsilon(u) = 0$, $\epsilon(1) = 1$ and antipode: $S_0(u) = -u$, $S_0(1) = 1$, for $u\in \mathfrak{g}$, and extending them by multiplicativity property to the entire $\U_{\mathfrak{g}}$. Recall that the universal enveloping algebra is a result of the factor construction
\begin{gather}\label{UEnvelop}
    \U_{\mathfrak{g}}=\frac{T\mathfrak{g}}{J_{\mathfrak{g}}},
\end{gather}
where $T\mathfrak{g}$ denotes tensor (free) algebra of the vector space $\mathfrak{g}$ quotient out by the ideal $J_\mathfrak{g}$ generated by elements $\langle X\otimes Y-Y\otimes X-[X, Y]\rangle$: $X, Y \in \mathfrak{g}$.
\end{example}

A (left) module algebra over a Hopf algebra $\H$ consist of a $\H$-module $\A$ which
is simulta\-neously an unital algebra satisfying the following compatibility
condition: 
\begin{gather}\label{genLeibniz}
L\triangleright (f\cdot g)=(L_{(1)}\triangleright f)\cdot (L_{(2)}\triangleright g)
\end{gather}
between multiplication $\cdot: \A\otimes\A\rightarrow \A$, coproduct $\Delta: \H\rightarrow\H\otimes\H$, $\Delta(L) =L_{(1)}\otimes L_{(2)}$, and (left) module action
$\triangleright:\H\otimes\A\rightarrow\A$; for $L\in\H$, $f,g\in\A$, $L\triangleright 1=\epsilon(L)$, $1\triangleright f=f$ (see, e.g., \cite{Klimyk, Kassel}).

In mathematics this condition plays a role of generalized Leibniz rule and it invokes exactly the Leibniz rule for primitive elements $\Delta(L)=L\otimes 1+1\otimes L$.
Instead, in physically oriented literature it is customary to call this condition a covariance condition and the corresponding algebra $\A$ a covariant quantum space (see, e.g., \cite{ABDMSW,DJMTWW,covQS}) with respect to $\H$. The covariance condition (\ref{genLeibniz}) entitles us also to introduce a new unital and associative algebra, the so-called smash (or crossed) product algebra $\A\rtimes\H$ \cite{Klimyk, Kassel,Majid2,smash}. Its structure is determined on the vector space $\A\otimes\H$ by multiplication:
\begin{gather}\label{smash}
(f\otimes L)\#(g\otimes M)=f(L_{(1)}\triangleright g)\otimes L_{(2)}M.
\end{gather}
Initial algebras are canonically embedded, $\A\ni f\shortrightarrow f\otimes 1$
and $\H\ni L\shortrightarrow 1\otimes L$~as subalgebras in $\A\rtimes\H$.\footnote{Further on, in order to simplify notation, we shall identify $(f\otimes 1)\#(1\otimes L)$ with $fL$, therefore~(\ref{smash}) rewrites simply as $(fL)\#(gM)=f(L_{(1)}\triangleright g)L_{(2)}M$.} Particularly, the trivial action
$L\triangleright g=\epsilon(L)g$ makes $\A\rtimes\H$ isomorphic to the ordinary tensor product algebra $\A\otimes\H$: $(f\otimes L)(g\otimes M)=fg\otimes LM$. It has a canonical Heisenberg representation on the vector space $\A$ which reads as follows:
\begin{gather*}
\hat{f}(g)= f\cdot g, \qquad \hat{L}(g)=L\triangleright g,
\end{gather*}
where $\hat{f}$, $\hat{L}$ are linear operators acting in $\A$, i.e.\ $\hat{f}, \hat{L}\in {\rm End}\, \A$. In other words around, the action~(\ref{genLeibniz}) extend to the action of entire $\A\rtimes\H$ on $\A$: $(f M)\triangleright g=f(M\triangleright g)$.

\begin{remark}\label{bicross}
 Note that $\A\rtimes\H$ is not a Hopf algebra in general. In fact, enactment of the Hopf algebra
 structure on $\A\rtimes\H$ involves some extras (e.g., co-action) and is related with the so-called bicrossproduct  construction~\cite{Majid2}. It has been shown in~\cite{MR}  that $\kappa$-Poincar\'{e} Hopf algeb\-ra~\cite{Luk1} admits bicrossproduct construction. We shall not concentrate on this topic here.
\end{remark}

\begin{example}\label{example2}
An interesting situation appears when the algebra $\A$ is a universal envelope of some Lie algebra $\mathfrak{h}$, i.e.\ $\A=\U_{\mathfrak{h}}$.
In this case it is enough to determine the Hopf action on generators $a_i\in \mathfrak{h}$ provided that the consistency conditions
\begin{gather}\label{smash2}
(L_{(1)}\triangleright a_i)\left( L_{(2)}\triangleright a_j\right) -(L_{(1)}\triangleright%
a_j)\left( L_{(2)}\triangleright a_i\right) -\imath c_{ij}^k L\triangleright a_k=0
\end{gather}
hold, where $[a_i, a_j]=\imath c_{ij}^ka_k$.
Clearly, (\ref{smash2}) allows to extend the action to entire algebra $\A$. For similar reasons the def\/inition of the action can be reduced to generators $L$ of $\H$.
\end{example}

\begin{example}\label{example3}
Another interesting case appears if $\H=\U_{\mathfrak{g}}$ for the Lie algebra
$\mathfrak{g}$ of some Lie group $G$ provided with the primitive Hopf algebra structure. This undergoes a ``geometrization'' procedures as follows. Assume one has given $G$-manifold $\M$. Thus $\mathfrak{g}$ acts via vector f\/ields on the (commutative) algebra of smooth functions $\A=C^\infty(\M)$.\footnote{In fact $\A$ can be chosen to be $\mathfrak{g}$-invariant subalgebra of $C^\infty(\M)$.} Therefore the Leibniz rule makes the compatibility conditions~(\ref{smash2}) warranted. The corresponding smash product algebra becomes an algebra of dif\/ferential operators on $\M$ with coef\/f\/icients in $\A$. A deformation of this geometric setting has been recently advocated as an alternative to quantization of gravity (see, e.g., \cite{ABDMSW,ADMW,Szabo2006} and references therein).
\end{example}

\begin{example}\label{example4}
A familiar Weyl algebra can be viewed as a crossed product of an algebra of translations $\mathfrak{T}^{n}$ containing $P_{\mu }$ generators with an algebra $\mathfrak{X}^n$ of spacetime coordinates $x^{\mu }$. More exactly, both algebras are def\/ined as a dual pair of the universal commutative algebras with $n$-generators (polynomial algebras), i.e.\ $\mathfrak{T}^{n}\equiv {\rm Poly}(P_\mu)\equiv\mathbb{C}[P_0,\ldots, P_{n-1}]$ and $\mathfrak{X}^{n}\equiv {\rm Poly}(x^\mu)\equiv\mathbb{C}[x^0,\ldots, x^{n-1}]$.\footnote{Here $n$  denotes a dimension of physical spacetime
which is not yet provided with any metric. Nevertheless, for the sake of future applications, we shall use ``relativistic'' notation with spacetime indices $\mu$ and $\nu $ running $0,\ldots,n-1$ and space indices $j,k=1,\ldots,n-1$ (see Example~\ref{ortho}).} Alternatively, both algebras are isomorphic to the universal enveloping algebra $\U_{\mathfrak{t}^{n}}\cong \mathfrak{T}^{n}\cong\mathfrak{X}^{n}$ of the $n$-dimensional Abelian Lie algebra $\mathfrak{t}^{n}$. Therefore one can make use of the primitive Hopf algebra structure on $\mathfrak{T}^{n}$ and extend the action implemented by duality map
\begin{gather}\label{actionP}
P_{\mu}\triangleright x^\nu=-\imath \langle P_{\mu}, x^\nu\rangle =-\imath\delta_\mu^\nu, \qquad
P_{\mu}\triangleright 1=0
\end{gather}
to whole algebra  $\mathfrak{X}^{n}$ due to the Leibniz rule, e.g.,
$P_{\mu}\triangleright (x^\nu x^\lambda)=-\imath\delta^\nu_\mu x^\lambda-\imath\delta^\lambda_\mu x^\nu$, induced by primitive coproduct $\Delta(P_\mu)=P_\mu\otimes 1+1\otimes P_\mu$, cf.~(\ref{smash}).
In result one obtains the following standard set of Weyl--Heisenberg commutation relations:
\begin{gather}\label{Weyl}
\left[ P_{\mu },x^{\nu }\right]_\#\equiv\left[ P_{\mu },x^{\nu }\right]=-\imath\delta _{\mu}^\nu\, 1, \qquad
\left[ x^{\mu },x^{\nu }\right]_\#\equiv\left[ x^{\mu },x^{\nu }\right]=
\left[ P_{\mu },P_{\nu }\right]_\#\equiv\left[ P_{\mu },P_{\nu }\right]=0.
\end{gather}
as generating relations\footnote{Hereafter we skip denoting commutators with $\#$ symbol when it is clear that they has been obtained by smash product construction.} for the Weyl algebra $\mathfrak{W}^{n}\equiv\mathfrak{X}^{n}\rtimes\mathfrak{T}^{n}$. In fact the algebra (\ref{Weyl}) represents the so-called Heisenberg double \cite{Klimyk,Kassel,Hdouble}. It means that $\mathfrak{T}^{n}$ and $\mathfrak{X}^{n}$ are dual pairs of Hopf algebras (with the primitive coproducts) and the action (\ref{actionP}) has the form: $P\triangleright x=\langle P, x_{(1)}\rangle x_{(2)}$.
In the Heisenberg representation $P_{\mu} \triangleright = -\imath\partial_\mu=-\imath\frac{\partial}{\partial x^\mu}$. For this reason  the
Weyl algebra is known as an algebra of dif\/ferential operators with polynomial coef\/f\/icients in $\mathbb{R}^n$.  \end{example}

\begin{remark}
The Weyl algebra as def\/ined above is not an enveloping algebra of any Lie algebra. It is due to the fact that the action (\ref{actionP}) is of ``$0$-order''. Therefore, it makes dif\/f\/icult to determine a Hopf algebra structure on it. The standard way to omit this problem relies on introducing the central element $C$ and replacing the commutation relations (\ref{Weyl}) by the following ones
\begin{gather}\label{Weyl--Heisenberg}
\left[ P_{\mu },x^{\nu }\right]=-\imath \delta _{\mu}^\nu C, \qquad
\left[ x^{\mu },x^{\nu }\right]=\left[C ,x^{\nu }\right]=
\left[ P_{\mu },P_{\nu }\right]=\left[C ,P_{\nu }\right]=0.
\end{gather}
The relations above determine $(2n+1)$-dimensional Heisenberg Lie algebra of rank $n+1$ . Thus Heisenberg algebra can be def\/ined as an enveloping algebra for (\ref{Weyl--Heisenberg}). We shall not follow this path here, however it may provide a starting point for Hopf algebraic deformations, see e.g.~\cite{LukMinn}.
\end{remark}

\begin{remark}
Real structure on a complex algebra can be def\/ined by an appropriate Hermitian conjugation~$\dag$. The most convenient way to introduce it, is by the indicating Hermitian gene\-ra\-tors. Thus the real structure corresponds to real algebras. For the Weyl algebra case, e.g., one can set the generators $(P_\mu, x^\nu)$ to be Hermitian, i.e.\ $(P_\mu)^\dag=P_\mu$, $(x^\nu)^\dag=x^\nu$. We shall not explore this point here. Nevertheless, all commutation relations below will be written in a form adopted to Hermitian realization. \end{remark}
\begin{remark}\label{Hspace}
Obviously, the Hopf action (\ref{actionP}) extends to the full algebra $C^\infty(\mathbb{R}^n)\otimes\mathbb{C}$ of complex valued smooth functions on $\mathbb{R}^n$. Its invariant subspace of compactly supported  functions $C_0^\infty(\mathbb{R}^n)\otimes\mathbb{C}$ form a dense domain in the Hilbert space of square-integrable functions: ${\cal L}^2(\mathbb{R}^n, dx^n)$. Consequently,  the Heisenberg  representation extends  to Hilbert space representation of $\mathfrak{W}^{n}$ by (unbounded) operators. This corresponds to canonical quantization procedure and in the relativistic case leads to St\"{u}ckelberg's version of Relativistic Quantum Mechanics~\cite{Menski}.
\end{remark}

\begin{example}\label{example5}
 Smash product generalizes also the notion of Lie algebra semidirect product. As an example one can consider a semidirect product of $\mathfrak{gl}(n)$ with the algebra of translations $\mathfrak{t}^n$: $\mathfrak{igl}(n)=\mathfrak{gl}(n)\niplus\mathfrak{t}^n$.  Thus $\U_{\mathfrak{igl}(n)}=\mathfrak{T}^{n}\rtimes\U_{\mathfrak{gl}(n)}$. Now the corresponding (left) Hopf action of $\mathfrak{gl}(n)$ 
on $\mathfrak{T}^n$ generators reads
\begin{gather*}
L^{\mu}_{\nu}\triangleright P_{\rho}=\imath\delta^\mu_\rho P_\nu.
\end{gather*}
The resulting algebra is described by a standard set of $\mathfrak{igl}(n)$ commutation relations:
\begin{gather}  \label{igl}
[L_{\nu }^{\mu },L_{\lambda }^{\rho }]=-\imath \delta _{\nu }^{\rho }L_{\lambda
}^{\mu }+\imath\delta _{\lambda }^{\mu }L_{\nu }^{\rho },\qquad    [L_{\nu
}^{\mu },P_{\lambda }]=\imath\delta _{\lambda }^{\mu }P_{\nu }, \qquad [P_\mu, P_\nu]=0.
\end{gather}

Analogously one can consider Weyl extension of $\mathfrak{gl}(n)$ as a
double crossed-product construction $\mathfrak{X}^n\rtimes(\mathfrak{T}^n\rtimes
\U_{\mathfrak{gl}(n)})$ with generating relations (\ref{igl})
supplemented by
\begin{gather*}
[L_{\nu}^{\mu }, x^{\lambda }]=-\imath\delta_{\nu }^{\lambda}x^{\mu } , \qquad
[P_\mu, x^\nu]=-\imath\delta_\mu^\nu , \qquad [x^\mu, x^\nu]=0.
\end{gather*}
The corresponding action is classical, i.e.\ it is implied by Heisenberg dif\/ferential representation (cf.\ formula~(\ref{Heisenberg2}) below):
\begin{gather}\label{action1}
 P_{\mu}\triangleright x^\nu=-\imath\delta_\mu^\nu, \qquad
   L^{\mu}_{\nu}\triangleright x^{\rho}=-\imath\delta_\nu^\rho x^\mu.
\end{gather}
Therefore, the Weyl algebra $\mathfrak{W}^n$ becomes a subalgebra in
$\mathfrak{X}^n\rtimes\left(\mathfrak{T}^n\rtimes\U_{\mathfrak{gl}(n)}\right)$.
Besides this isomorphic embedding one has a surjective algebra homomorphism
$\mathfrak{X}^n\rtimes\left(\mathfrak{T}^n\rtimes\U_{\mathfrak{gl}(n)}\right)\rightarrow\mathfrak{W}^n$
provided by
\begin{gather}\label{Heisenberg1a}
P_\mu \rightarrow P_\mu , \qquad
x^\mu\rightarrow x^\mu , \qquad L^\nu_\mu \rightarrow x^\nu P_\mu.
\end{gather}
We shall call this epimorphism Weyl algebra (or Heisenberg) realization of
$\mathfrak{X}^n\rtimes\left(\mathfrak{T}^n\rtimes\U_{\mathfrak{gl}(n)}\right)$.
Particularly, the map $L^\nu_\mu \rightarrow x^\nu P_\nu$ is a Lie algebra isomorphism.
The Heisenberg realization described above induces Heisenberg representation of $\mathfrak{X}^n\rtimes\left(\mathfrak{T}^n\rtimes\U_{\mathfrak{gl}(n)}\right)$
\begin{gather}\label{Heisenberg2} P_\mu\,\triangleright=-\imath\partial_\mu\equiv-\imath\frac{\partial}{\partial x^\mu}, \qquad
x^\mu=x^\mu ,\qquad L^\nu_\mu
\triangleright=-\imath x^\nu\partial_\mu
\end{gather}
acting in the vector space $\mathfrak{X}^n$. One can notice that (\ref{Heisenberg2}) can be extended to the vector space $C^\infty(\mathbb{R}^n)\otimes\mathbb{C}$ and f\/inally to the Hilbert space representation in ${\cal L}^2(\mathbb{R}^n, dx^n)$.
\end{example}
More examples one can f\/ind noticing that the Lie algebra $\mathfrak{igl}(n)$ contains several interesting subalgebras, e.g., inhomogeneous special linear transformations $\mathfrak{%
isl}(n)=\mathfrak{sl}(n)\niplus\mathfrak{t}^{n}$.

\begin{example}\label{ortho}
Even more interesting and important for physical applications example is provided by the inhomogeneous orthogonal transformations $\mathfrak{io}(g; n)=\mathfrak{o}%
(g; n)\niplus\mathfrak{t}^{n}\subset \mathfrak{isl}(n)$,  which Weyl extension is def\/ined by the following set of commutation relations:
\begin{gather}  \label{isog1}
[M_{\mu \nu },M_{\rho \lambda }]= \imath g_{\mu \rho }M_{\nu
\lambda }-\imath g_{\nu \rho }M_{\mu \lambda }- \imath g_{\mu
\lambda }M_{\nu \rho }
+\imath g_{\nu \lambda }M_{\mu \rho },
\\  \label{isog2}
[M_{\mu \nu },P_{\lambda }]= \imath g_{\mu \lambda}P_{\nu }-\imath g_{\nu \lambda }P_{\mu }
, \qquad
[M_{\mu \nu },x_{\lambda }]= \imath g_{\mu \lambda}x_{\nu }-\imath g_{\nu \lambda }x_{\mu },
\\
\label{Weyl_g}
\left[ P_{\mu },x_{\nu }\right]= -\imath g_{\mu\nu}, \qquad
\left[ x_{\mu },x_{\nu }\right]=
\left[ P_{\mu },P_{\nu }\right]=0,
\end{gather}
where
\begin{gather}  \label{isog3}
M_{\mu \nu }=g_{\mu \lambda }L_{\nu }^{\lambda }-g_{\nu \lambda }L_{\mu
}^{\lambda }.
\end{gather}
are def\/ined by means of the (pseudo-Euclidean)\footnote{We shall write $\mathfrak{io}(n-p, p)$ whenever the signature $p$ of the metric will become important. In fact, dif\/ferent metric's signatures lead to dif\/ferent real forms of $\mathfrak{io}(n,\mathbb{C})$.} metric tensor $g_{\mu\nu}$ and $x_\lambda=g_{\lambda\nu}x^\nu$.
The last formula determines together with (\ref{action1}) the classical action of $\mathfrak{io}(n,\mathbb{C})$ on  $\mathfrak{X}^n$.
Thus relations (\ref{isog1})--(\ref{Weyl_g}) determine Weyl extension of the inhomogeneous orthogonal Lie algebra $\mathfrak{X}^n\rtimes(\mathfrak{T}^n\rtimes\U_{\mathfrak{o}(g; n)})$ as subalgebra
in $\mathfrak{X}^n\rtimes(\mathfrak{T}^n\rtimes\U_{\mathfrak{gl}(n)})$. In particular, the case of Poincar\'{e} Lie algebra $\mathfrak{io}(1,3)$ will be studied in more details in Section~\ref{section4}.
\end{example}

\begin{remark}\label{StablForm} Quantum groups include Hopf-algebraic deformations of Lie algebras.
There exists also purely Lie-algebraic framework for deformations of Lie algebras. Accordingly,
all semisimple Lie algebras are stable.
The algebra (\ref{isog1})--(\ref{Weyl_g}) has been investigated in~\cite{Mendes} from the point of view of the stability problem (see also~\cite{Chryss}). It was shown that for the Lorentzian signature of $g_{\mu\nu}$ stable forms of (\ref{isog1})--(\ref{Weyl_g}) are provided by simple Lie algebras: $\mathfrak{o}(3, n-1)$, $\mathfrak{o}(2, n)$ or $\mathfrak{o}(1, n+1)$. Relationships between Lie and Hopf algebraic frameworks for deformations have been discussed in~\cite{Kassel}.
\end{remark}

\section[$\kappa$-Minkowski spacetime by Drinfeld twist]{$\boldsymbol{\kappa}$-Minkowski spacetime by Drinfeld twist}\label{section3}

Having the Hopf-module algebra structure one can use the deformation theory to construct new (deformed) objects which still possess the same structure.
Assume a Hopf module algebra $\A$ over $\mathcal{H}$.
These can be deformed, by a suitable twisting element $\mathcal{F}$,
to achieve the deformed Hopf module algebra $(\mathcal{A}^{\mathcal{F}},\mathcal{H}^{\mathcal{F}})$, where the algebra $\mathcal{A}^{\mathcal{F}}$ is equipped with a twisted (deformed) star-product (see~\cite{Oeckl,ABDMSW} and references therein)
\begin{gather}  \label{tsp}
x\star y=m \circ\mathcal{F}^{-1}\triangleright (x\otimes y)=(\bar{\mathrm{f}}^\alpha\triangleright x)\cdot(\bar{%
\mathrm{f}}_\alpha\triangleright y)
\end{gather}
while the Hopf action $\triangleright$ remains unchanged.
Hereafter the twisting element $\mathcal{F}$ is symbolically written in the
following form:
\begin{gather*}
  \mathcal{F}=\mathrm{f}^\alpha\otimes\mathrm{f}_\alpha\in \H\otimes\H \qquad \mbox{and}
\qquad \mathcal{F}^{-1}=\bar{\mathrm{f}}^\alpha\otimes\bar{\mathrm{f}}_\alpha\in \H\otimes\H
\end{gather*}
and belongs to the Hopf algebra $\H$. The corresponding smash product $\mathcal{A}^{\mathcal{F}}\rtimes\mathcal{H}^{\mathcal{F}}$
has deformed cross-commutation relations (\ref{smash}) determined by deformed coproduct $\Delta^{\mathcal{F}}$.
Before proceeding further let us remind in more details that the quantized Hopf algebra $\H^{\mathcal{F}}$ has non-deformed algebraic sector (commutators),
while coproducts and antipodes are subject of the deformation:
\begin{gather*}
\Delta^{\mathcal{F}} (\cdot)=\mathcal{F} \Delta (\cdot)\mathcal{F}^{-1}
,\qquad S^{\mathcal{F}}(\cdot)=u S(\cdot) u^{-1},
\end{gather*}
where $u=\mathrm{f}^\alpha S(\mathrm{f}_\alpha)$.
The twisting two-tensor $\mathcal{F}$ is an invertible element in $\H\otimes \H$ which fulf\/ills the 2-cocycle and normalization
conditions \cite{Drinfeld,Chiari}:
\begin{gather*}  
 \mathcal{F}_{12}(\Delta\otimes {\rm id})\mathcal{F}=\mathcal{F}
_{23}({\rm id}\otimes\Delta)\mathcal{F} ,\qquad   (\epsilon\otimes {\rm id})\mathcal{F}
=1=({\rm id}\otimes\epsilon)\mathcal{F},
\end{gather*}
which guarantee co-associativity of the deformed coproduct $\Delta^{\mathcal{F}}$ and associativity of the corresponding twisted star-product~(\ref{tsp}). Moreover, it implies simultaneously that deformed and undeformed smash product are isomorphic:

\begin{proposition}\label{prop1} For any Drinfeld twist $\mathcal{F}$  the twisted smash product
algebra  $\mathcal{A}^{\mathcal{F}}\rtimes\mathcal{H}^{\mathcal{F}}$ is  isomorphic to the initial $($undeformed$)$ one $\mathcal{A}\rtimes\mathcal{H}$. In other words the algebra $\mathcal{A}^{\mathcal{F}}\rtimes\mathcal{H}^{\mathcal{F}}$ is twist independent and can be realized by a change of generators in the algebra $\mathcal{A}\rtimes\mathcal{H}$.
\end{proposition}

Notice that subalgebras $\mathcal{A}$ and $\mathcal{A}^{\mathcal{F}}$ are not isomorphic. In the case of commutative $\mathcal{A}$ we use to call noncommutative algebra $\mathcal{A}^{\mathcal{F}}$ quantization of $\mathcal{A}$. Similarly $\mathcal{H}$ and $\mathcal{H}^{\mathcal{F}}$ are not isomorphic as Hopf algebras.

\begin{proof}\label{proof1}
First of all we notice that the inverse twist $\mathcal{F}^{-1}=\bar{\mathrm{f}}^\alpha\otimes\bar{\mathrm{f}}_\alpha$ satisf\/ies analogical
cocycle condition $(\Delta\otimes {\rm id})(\mathcal{F}^{-1})\mathcal{F}^{-1}_{12}=({\rm id}\otimes\Delta)(\mathcal{F}^{-1})\mathcal{F}_{23}^{-1}$.
It reads as
\[
\bar{\mathrm{f}}^\alpha_{(1)}\bar{\mathrm{f}}^\beta\otimes \bar{\mathrm{f}}^\alpha_{(2)}\bar{\mathrm{f}}_\beta\otimes\bar{\mathrm{f}}_\alpha=
\bar{\mathrm{f}}^\alpha\otimes\bar{\mathrm{f}}_{\alpha (1)} \bar{\mathrm{f}}^\beta\otimes\bar{\mathrm{f}}_{\alpha (2)}\bar{\mathrm{f}}_\beta.
\]
For any element $x\in \mathcal{A}$ one can associate the corresponding element
$\hat x=(\mathrm{\bar{f}}^\alpha\triangleright x)\cdot\bar{\mathrm{f}}_\alpha\in \mathcal{A}\rtimes\mathcal{H}$. Thus~(\ref{tsp}) together with the cocycle condition gives $\hat x\cdot\hat y=(\bar{\mathrm{f}}^\alpha\triangleright (x\star y))\cdot\bar{\mathrm{f}}_\alpha$.
Due to invertibility of twist we can express the original elements $x$ as a functions of the deformed one:
$x=(\mathrm{f}^\alpha\triangleright \hat x)\star\mathrm{f}_\alpha $.
It means that subalgebra generated by the elements $\hat x$ is isomorphic to $\mathcal{A}^{\mathcal{F}}$.
Notice that $\mathcal{A}=\mathcal{A}^{\mathcal{F}}$, $\mathcal{H}=\mathcal{H}^{\mathcal{F}}$
and $\mathcal{A}\rtimes\mathcal{H}=\mathcal{A}^{\mathcal{F}}\rtimes\mathcal{H}^{\mathcal{F}}$
as linear spaces.

The requested isomorphism $\varphi: \mathcal{A}^{\mathcal{F}}\rtimes\mathcal{H}^{\mathcal{F}}\rightarrow\mathcal{A}\rtimes\mathcal{H}$
can be now def\/ined by an invertible mapping
\begin{gather}\label{change}
\mathcal{A}^{\mathcal{F}}\ni x\rightarrow \varphi(x)=\hat x \in\mathcal{A}\rtimes\mathcal{H}\qquad
\hbox{and}\qquad \mathcal{H}^{\mathcal{F}}\ni L\rightarrow \varphi(L)=L\in \mathcal{A}\rtimes\mathcal{H}.
\end{gather}
Utilizing again the cocycle condition one checks that $\varphi(L\star x)=L\cdot\hat x=(L_{(1)}\mathrm{\bar{f}}^\alpha\triangleright x)\cdot L_{(2)}\bar{\mathrm{f}}_\alpha $.

Let $(x^\mu)$ be a set of generators for $\mathcal{A}$ and  $(L^k)$ be a set of generators for $\mathcal{H}$. Then the isomorphism (\ref{change}) can be described as a change of generators (``basis''):
$(x^\mu, L^k)\rightarrow (\hat x^\mu, L^k)$ in $\mathcal{A}\rtimes\mathcal{H}$.
\end{proof}
Dif\/ferent scheme for ``unbraiding of braided tensor product'' has been presented in~\cite{Fiore1}.

\begin{remark}\label{trivialTwist}
Consider internal automorphism of the algebra  $\mathcal{H}$ given by the similarity transformation $L\rightarrow WLW^{-1}$, where $W$ is some invertible element in $\mathcal{H}$.
This automorphism induces the corresponding isomorphism of Hopf algebras, which can be equivalently described as coalgebra isomorphism $(\mathcal{H}, \Delta)\rightarrow (\mathcal{H}, \Delta^{\mathcal{F}_W})$, where $\Delta^{\mathcal{F}_W}$ denotes twisted coproduct. Here $\mathcal{F}_W=W^{-1}\otimes W^{-1}\Delta(W)$ denotes the so called trivial (coboundary) twist. Of course, the twisted module algebra $\mathcal{A}^{\mathcal{F}_W}$ becomes isomorphic to the undeformed $\mathcal{A}$. Substituting
$W=\exp u$ one gets explicit form of the twisting element $\mathcal{F}_W=\exp{(-u\otimes 1-1\otimes u)}\exp{\Delta(u)}=\exp{(-\Delta_0(u))}\exp{\Delta(u)}$.
\end{remark}

As it is well-known from the general framework \cite{Drinfeld}, a twisted deformation of Lie algebra $\mathfrak{g}$ requires a topological extension of the corresponding enveloping algebra $\U_{\mathfrak{g}}$ into an algebra of formal power series $%
\U_{\mathfrak{g}}[[h]]$ in the formal parameter $h$ (see Appendix and, e.g.,~\cite{Klimyk,Kassel,Chiari,Bonneau} for deeper exposition)\footnote{This is mainly due to the fact that twisting element has to be invertible.}.
This provides the so-called $h$-adic topology. There is a correspondence between twisting element, which can be now  rewritten as a power series expansion
\[
\mathcal{F}=1\otimes 1+
\sum_{m=1}^\infty h^m\, \mathrm{f}^{(m)}\otimes\mathrm{f}_{(m)}\qquad\hbox{and}\qquad
\mathcal{F}^{-1}=1\otimes 1+
\sum_{m=1}^\infty h^m\, \bar{\mathrm{f}}^{(m)}\otimes\bar{\mathrm{f}}_{(m)},
\]
$\mathrm{f}^{(m)}, \mathrm{f}_{(m)}, \bar{\mathrm{f}}^{(m)}, \bar{\mathrm{f}}_{(m)}\in \U_{\mathfrak{g}}$, classical $r$-matrix $\mathfrak{r}\in \mathfrak{g}\wedge\mathfrak{g}$ satisfying  classical Yang--Baxter equation and universal (quantum) $r$-matrix $\mathcal{R}$:
\begin{gather*} 
\mathcal{R}=\mathcal{F}^{21}\mathcal{F}^{-1}=1+h \mathfrak{r}
\mod \big(h^2\big)
\end{gather*}
satisfying quantum Yang--Baxter equation. Moreover, classical $r$-matrices classify non-equivalent deformations.
Accordingly, the Hopf module algebra $\mathcal{A}$ has to be extended by $h$-adic topology to $\mathcal{A}[[h]]$ as well\footnote{The operator algebra setting for quantum groups and quantum spaces is admittedly much more heavy. Unfortunately, the
passage from quantized Lie algebra level to the function algebra level is not very straightforward and sometime even not possible, see e.g.~\cite{Op}.}.

\begin{remark}\label{Twist}
In general, there is no constructive way to obtain twist for a given classical $r$-mat\-rix. Few examples are known in the literature, e.g., Abelian~\cite{Reshetikhin}, Jordanian~\cite{Ogievetsky}  extended Jordanian twists \cite{Kulish} (see also \cite{Bonneau,VNT1}) as well as some of their combinations~\cite{Varna}. Twisted deformations of relativistic symmetries have been studied for a long time, see, e.g., \cite{BLT2} and references therein.
Almost complete classif\/ication of classical $r$-matrices for Lorentz an Poincar\'{e} group
has been given in~\cite{Zakrzewski} (see also~\cite{Lya}).
\end{remark}

\begin{example}\label{example6}
Let us consider the twist deformation of the algebra from Example~\ref{example5}. $h$-adic extension $\U_{\mathfrak{igl}(n)}\shortrightarrow\U_{\mathfrak{igl}(n)}[[h]]$ forces us to extend polynomial algebra:
$\mathfrak{X}^{n}\shortrightarrow\mathfrak{X}^{n}[[h]]$, which remains to be (undeformed) module algebra under the $\mathbb{C}[[h]]$-extended Hopf action $\triangleright$. Therefore their smash product contains $h$-adic extension of the Weyl algebra: $\mathfrak{W}^{n}[[h]]=\mathfrak{X}^{n}[[h]]\rtimes\mathfrak{T}^{n}[[h]]\subset \mathfrak{X}^{n}[[h]]\rtimes\left(\mathfrak{T}^{n}[[h]]\rtimes \U_{\mathfrak{gl}(n)}[[h]]\right)$.
After  having done $\mathfrak{igl}(n)$-twist $\mathcal{F}$ we can now deform simultaneously  both structures:  $U_{\mathfrak{igl}(n)}[[h]]\mapsto U_{\mathfrak{igl}(n)}[[h]]^{\mathcal{F}}$ and $\mathfrak{X}^{n}[[h]]\mapsto(\mathfrak{X}^{n}[[h]])^{\mathcal{F}}$ keeping the Hopf action $\triangleright$ unchanged. Deformed algebra $\mathfrak{X}^n[[h]]^\mathcal{F}$ has deformed star multipication $\star$ and can be represented by deformed $\star$-commutation relations
\begin{gather}\label{tcr}
  [x^{\mu },x^{\nu }]_{\star }\equiv x^{\mu }\star x^{\nu }-x^{\nu
}\star x^{\mu }=\imath\,h \theta^{\mu\nu}(x)\equiv
\imath h\big(\theta^{\mu\nu}+\theta^{\mu\nu}_\lambda x^\lambda+\theta^{\mu\nu}_{\lambda\rho} x^\lambda x^\rho+\cdots\big)
\end{gather}
replacing the undeformed (commutative) one
\begin{gather*}
[x^{\mu },x^{\nu }]=0,
\end{gather*}
where the coordinate functions $(x^\mu)$ play a role of generators for the
corresponding algebras:  deformed and undeformed one.  We will see on the examples below that many dif\/ferent twisted star products may lead to the same commutation relations~(\ref{tcr}). Particularly, one can def\/ine deformed Weyl algebra
$\mathfrak{W}^{n}[[h]]^\mathcal{F}=\mathfrak{X}^{n}[[h]]^\mathcal{F}
\rtimes\mathfrak{T}^{n}[[h]]^\mathcal{F}$, where $\mathfrak{T}^{n}[[h]]^\mathcal{F}$
denotes the corresponding Hopf subalgebra of deformed momenta in $\left(\mathfrak{T}^{n}[[h]]\rtimes \U_{\mathfrak{gl}(n)}[[h]]\right)^\mathcal{F}$.
\end{example}

\begin{remark}\label{Poisson}
The deformed algebra $\mathfrak{X}^{n}[[h]]^\mathcal{F}$ provides a deformation quantization of $\mathbb{R}^n$ equip\-ped with the Poisson structure (brackets)~\cite{Kon,BFFLS}
\begin{gather}\label{Pstr}
\{x^{\mu },x^{\nu }\}=
\theta^{\mu\nu}(x)\equiv
\theta^{\mu\nu}+\theta^{\mu\nu}_\lambda x^\lambda+\theta^{\mu\nu}_{\lambda\rho}x^\lambda x^\rho+\cdots,
\end{gather}
represented by Poisson bivector $\theta=\theta^{\mu\nu}(x)\partial_\mu\wedge\partial_\nu$. It is assumed that $\theta^{\mu\nu}(x)$ are polynomial functions, i.e.\ the sum in (\ref{Pstr}) is f\/inite where $\theta^{\mu\nu},\theta^{\mu\nu}_\lambda,\theta^{\mu\nu}_{\lambda\rho},\ldots$ are real numbers.
\end{remark}
\begin{proposition}\label{prop2}
Proposition {\rm \ref{prop1}} implies that $ \mathfrak{X}^{n}[[h]]\rtimes \U_{\mathfrak{igl}(n)}[[h]]$ is $\mathbb{C}[[h]]$ isomorphic to $ \mathfrak{X}^{n}[[h]]^\mathcal{F}\rtimes( \U_{\mathfrak{igl}(n)}[[h]])^\mathcal{F}$. Moreover, this isomorphism is congruent to the identity map mo\-dulo~$h$.
\end{proposition}

Replacing elements $\overline{\mathrm{f}}^{(m)}$,
$\overline{\mathrm{f}}_{(m)}\in\U_{\mathfrak{igl}(n)}$
in the formulae
\begin{gather}\label{Heis3}
\hat{x}^\mu=x^\mu+\sum_{m=1}^\infty h^m  \big(\overline{\mathrm{f}}^{(m)}
\triangleright x^\mu\big)\cdot\overline{\mathrm{f}}_{(m)}
\end{gather}
from Proposition~\ref{prop1}
by using Heisenberg realizations (\ref{Heisenberg1a}) one gets
\begin{proposition}\label{prop3}
All $\mathfrak{igl}(n)$-twist deformed Weyl algebras $\mathfrak{W}^{n}[[h]]^\mathcal{F}$
are $\mathbb{C}[[h]]$-isomorphic to undeformed $h$-adic extended Weyl algebra $\mathfrak{W}^{n}[[h]]$.
In this sense we can say that $\mathfrak{W}^{n}[[h]]^\mathcal{F}$
is a~pseudo-deformation of  $\mathfrak{W}^{n}[[h]]$ since the latter one
can be obtained by $($nonlinear and invertible$)$ change of generators from the first one\footnote{Cf.~\cite{Fiore2} for dif\/ferent approach to deformation of Clif\/ford and Weyl algebras.}.
\end{proposition}

\begin{remark}\label{h-repr}
Replacing again elements $\overline{\mathrm{f}}^{(m)}, \overline{\mathrm{f}}_{(m)}\in\U_{\mathfrak{igl}(n)}$ in (\ref{Heis3}) by their Heisenberg
representation: $P_\mu=-\imath\partial_\mu$ (cf.\ Example \ref{example5}) one obtains Heisenberg representation of $\mathfrak{W}^{n}[[h]]^\mathcal{F}$ and entire algebra $ \mathfrak{X}^{n}[[h]]^\mathcal{F}\rtimes( \U_{\mathfrak{igl}(n)}[[h]])^\mathcal{F}$ in the vector space $\mathfrak{X}^{n}[[h]]$.
Moreover, one can extend  Hilbert space representation from Remark \ref{Hspace} to the representation acting in ${\cal L}^2(\mathbb{R}^n, dx^n)[[h]]$. This is due to general theory of representations for $h$-adic quantum groups, see, e.g., \cite{Chiari}.
\end{remark}
Below we shall present explicitly  two one-parameter families of twists
corresponding to twisted star-product realization  of the well-known $\kappa $-deformed Minkowski spacetime algebra \cite{MR,Z}. Here $(\mathfrak{X}^{n}[[h]])^{\mathcal{F}}$ is generated by the relations:
\begin{gather}\label{kM1}
[x^0,x^k]_\star=\imath h x^k,\qquad [x^k,x^j]_\star=0  ,\qquad k,j=1,\ldots ,n-1,
\end{gather}
and constitutes a covariant algebra over deformed $\mathfrak{igl}(n)$.\footnote{Notice that $h$ is the formal parameter. So (\ref{kM1}) is not, strictly speaking, a Lie algebra.} Although the result of star multiplication of two generators  $x^\mu\star x^\nu$ is explicitly twist dependent, the generating relations (\ref{kM1}) are twist independent. Therefore all algebras
$(\mathfrak{X}^{n}[[h]])^{\mathcal{F}}$ are mutually isomorphic to each other. They provide a covariant deformation quantization of the $\kappa$-Minkowski Poisson structure represented by the linear Poisson bivector
\begin{gather}\label{kM2}
    \theta_{\kappa M}=x^k\partial_k\wedge \partial_0,\qquad k=1,\ldots, n-1
\end{gather}
on $\mathbb{R}^{n}$ (cf.~\cite{Z}). The corresponding Poisson tensor $\theta^{\mu\nu}(x)=a^\mu x^\nu - a^\nu x^\mu$, where $a^\mu=(1,0,0,0)$ is degenerated ($\det[\theta^{\mu\nu}(x)]=0$), therefore, the associated symplectic form $[\theta^{\mu\nu}(x)]^{-1}$ does not exist.
\begin{remark}\label{Poisson2}
More generally, there is one-to-one correspondence between linear Poisson structures
$\theta=\theta^{\mu\nu}_\lambda x^\lambda\partial_\mu\wedge\partial_\nu$ on $\mathbb{R}^{n}$ and $n$-dimensional Lie algebras $\mathfrak{g}\equiv\mathfrak{g}(\theta)$
\begin{gather}\label{envelop0}
    [X^\mu, X^\nu]=\imath \theta^{\mu\nu}_\lambda X^\lambda
\end{gather}
with the constants $\theta^{\mu\nu}_\lambda$ playing the role of Lie algebra structure constants.
Therefore, they are also called  Lie--Poisson structures. Following Kontsevich we shall call the corresponding universal enveloping algebra $\U_{\mathfrak{g}}$ a canonical quantization of $(\mathfrak{g}^*,  \theta)$~\cite{Kon}. The classif\/ication of  Lie--Poisson structures on
$\mathbb{R}^{4}$ has been recently presented  in~\cite{Sheng}.

Let us consider the following modif\/ication of the universal enveloping algebra construction~(\ref{UEnvelop})
\begin{gather*}
    \U_h(\mathfrak{g})=\frac{T\mathfrak{g}[[h]]}{J_{h}},
\end{gather*}
where $J_{h}$ denotes an ideal generated by elements
$\langle X\otimes Y-Y\otimes X-h[X, Y]\rangle $ and is closed in $h$-adic topology. In other words the algebra $\U_h(\mathfrak{g})$ is $h$-adic unital algebra generated by $h$-shifted relations
\begin{gather*}
    [\X^\mu, \X^\nu]=\imath h\theta^{\mu\nu}_\lambda \X^\lambda
\end{gather*}
imitating  the Lie algebraic ones~(\ref{envelop0}).
This algebra provides the so-called universal quantization of $(\mathfrak{X}^{n}, \theta)$~\cite{Kat}.
Moreover, due to universal (quotient) construction there is a $\mathbb{C}[[h]]$-algebra epimorphism from $\U_h(\mathfrak{g})$ onto $(\mathfrak{X}^{n}[[h]])^{\mathcal{F}}$ for a suitable twist $\mathcal{F}$ (cf.~(\ref{tcr})). In fact, $\U_h(\mathfrak{g})$ can be identif\/ied with $\mathbb{C}[[h]]$-subalgebra in $\U_{\mathfrak{g}}[[h]]$ generated by $h$-shifted generators $\X^\mu=hX^\mu$.  $\U_{\mathfrak{g}}[[h]]$~is by construction a topological free $\mathbb{C}[[h]]$-module (cf.~Appendix). For the case of  $\U_h(\mathfrak{g})$ this question is open.
\end{remark}

\begin{example}\label{example8}
In the the case of $\theta_{\kappa M}$ (\ref{kM2}) the corresponding Lie algebra is a solvable one. Following~\cite{Oriti} we shall denote it as $\mathfrak{an}^{n-1}$. The universal $\kappa$-Minkowski  spacetime algebra $\U_h(\mathfrak{an}^{n-1})$ has been introduced in~\cite{Z}, while its Lie algebraic counterpart the canonical $\kappa$-Minkowski  spacetime algebra $\U_{\mathfrak{an}^{n-1}}$ in~\cite{MR}. We shall consider both algebras in more details later on.
\end{example}

\subsection*{Abelian family of twists providing $\boldsymbol{\kappa}$-Minkowski spacetime}

The simplest possible twist is Abelian one.
$\kappa $-Minkowski spacetime can be  implemented by the one-parameter
family of Abelian twists \cite{BP} with $s$ being a numerical
parameter labeling dif\/ferent twisting tensors (see also~\cite{Bu, MSSKG}):
\begin{gather}  \label{At}
  \mathfrak{A}_{s}=\exp \left[ \imath h \left(sP_{0}\otimes
D-\left( 1-s\right) D\otimes
P_{0}\right)\right],
\end{gather}
where $D=\sum\limits_{\mu=0}^{n-1}L^\mu_\mu-L^0_0$.
All twists correspond to the same classical $r$-matrix\footnote{Since in the Heisenberg realization the space dilatation generator $D=x^k\partial_k$, $\mathfrak{r}$ coincides with the Poisson bivector~(\ref{kM2}).}:
\begin{gather*}
\mathfrak{r}=D\wedge P_0
\end{gather*}
and they have the same universal quantum r-matrix which is  of exponential form:
\begin{gather*}
\mathcal{R}=\mathfrak{A}_s^{21}\mathfrak{A}_s^{-1}=e^{\imath D\wedge P_0}.
\end{gather*}

\begin{remark}
This implies that corresponding Hopf algebra deformations of $\U_{\mathfrak{igl}(n)}^\mathcal{F}$ for dif\/ferent values of parameter $s$ are isomorphic. Indeed, $\mathfrak{A}_{s}$ are related by trivial twist:  $\mathfrak{A}_{s_2}=\mathfrak{A}_{s_1}{\mathcal F}_{W_{12}}$, where $W_{12}=\exp{(\imath(s_1-s_2) a DP_0)}$ (cf.\ Remark~(\ref{trivialTwist})). Consequently $ \Delta^{\rm op}_{(s=0)}=\Delta_{(s=1)}$ and
$\Delta^{\rm op}=\mathcal{R}\Delta\mathcal{R}^{-1}$.
We will see later on that dif\/ferent values of $s$ lead to dif\/ferent Heisenberg realizations and describe dif\/ferent physical models (cf.~\cite{BP}).
\end{remark}

\begin{remark}
The smallest subalgebra generated by $D$, $P_0$ and the Lorentz generators (\ref{isog3}) turns out to be entire $\mathfrak{igl}(n)$ algebra.
Therefore, as a consequence,  the deformation induced by twists (\ref{At}) cannot be restricted to
the inhomogeneous orthogonal transformations (\ref{isog1})--(\ref{isog2}). One can say that Abelian twists~(\ref{At}) are genuine $\mathfrak{igl}(n)$-twists.
\end{remark}

The deformed coproducts read as follows (cf.~\cite{Bu,BP}):
\begin{gather*}
\Delta _{s}\left( P_{0}\right) =1\otimes P_{0}+P_{0}\otimes 1,
\qquad
\Delta _{s}\left( P_{k}\right) =e^{-hsP_{0}}\otimes P_{k}+P_{k}\otimes
e^{h(1-s)P_{0}},
\\
\Delta _{s}\left( L_{k}^{m}\right) =1\otimes L_{k}^{m}+L_{k}^{m}\otimes 1,
\qquad
\Delta _{s}\left( L_{0}^{k}\right) =e^{hsP_{0}}\otimes
L_{0}^{k}+L_{0}^{k}\otimes e^{-h(1-s)P_{0}},
\\
\Delta _{s}\left( L_{k}^{0}\right) =e^{-hsP_{0}}\otimes
L_{k}^{0}+L_{k}^{0}\otimes e^{h(1-s)P_{0}}+hsP_{k}\otimes De^{h(1-s)P_{0}}
-h(1-s)D\otimes P_{k},
\\
\Delta _{s}\left( L_{0}^{0}\right) =1\otimes L_{0}^{0}+L_{0}^{0}\otimes 1
+hsP_{0}\otimes D-h(1-s)D\otimes P_{0},
\end{gather*}
and the antipodes are:
\begin{gather*}
S_s\left( P_{0}\right) =-P_{0}, \qquad S_s\left( P_{k }\right) =-P_{k
}e^{hP_0},
\\
S_s\left( L_{k}^{m}\right) =-L_{k}^{m}, \qquad S_s\left( L_{0}^{k}\right)
=-L_{0}^{k} e^{-hP_0},
\qquad
S_s\left( L_{0}^{0}\right) =-L_{0}^{0}-h(1-2s)DP_0,
\\
S_s\left( L_{k}^{0}\right) =-e^{hsP_0}L_{k}^{0}e^{-
h(1-s)P_0}+h\big[sP_kDe^{hsP_0}+(1-s)DP_k e^{h(1+s)P_0}\big].
\end{gather*}
The above relations are particularly simple for $s=0, 1, {\frac{1}{2}}$.
Smash product construction (for given $\Delta_s$) together with classical action~(\ref{action1}) leads to the following crossed commutators:
\begin{gather*}
\left[ \hat{x}^{\mu },P_{0}\right]_s =\imath \delta^\mu_0,\qquad\left[ \hat{x}^\mu,P_{k}\right]_s =\imath \delta^\mu_ke^{h(1-s)P_{0}}-\imath hs\delta^\mu_0P_{k},
\\
\left[ \hat{x}^\mu,L_{k}^{m}\right]_s =\imath \delta^\mu_k\hat{x}^{m},\qquad \big[ \hat{x}^\mu,L_{0}^{k}\big]_s =\imath \delta^\mu_0\hat{x}^{k}e^{-h(1-s)P_{0}}-\imath hs\delta^\mu_0L_{0}^{k},
\\
\left[ \hat{x}^{\mu }, L_{k}^{0}\right]_s =\imath
\delta_k^\mu\hat{x}^{0} e^{h(1-s)P_{0}}-\imath hs\delta^\mu_0L_{k}^{0}
+\imath hs\delta_k^\mu D-\imath h(1-s)\delta^\mu_l \hat{x}^{l}P_{k},
\\
\left[\hat{x}^{\mu },L_{0}^{0}\right]_s =\imath \delta_0^\mu\hat{x}^{0}
+\imath hs\delta_{0}^{\mu }D-\imath h(1-s)\delta_k^\mu \hat{x}^kP_{0}.
\end{gather*}
Supplemented with (\ref{igl}) and $\kappa$-Minkowski spacetime commutators:
\begin{gather}\label{kMhat}
  [\hat{x}^0,\hat{x}^k]=\imath h \hat{x}^k, \qquad [\hat{x}^k,\hat{x}^j]=0  ,\qquad k,j=1,\ldots, n-1
\end{gather} they form the algebra:
$\mathfrak{X}^n[[h]]^\mathcal{F}\rtimes \U_{\mathfrak{igl}(n)}^\mathcal{F}$.
 The change of generators $(L_\mu^\nu,P_\rho,x^\lambda)\shortrightarrow(L_\mu^\nu,P_\rho,\hat{x}^\lambda_s)$, where
 \begin{gather*}  
\hat x^{i}_{s}=x^{i}e^{\left( 1-s\right)hP_0} , \qquad \hat x^{0}_{s} =
x^{0}- h s D
\end{gather*}
 implies the isomorphism from Proposition~\ref{prop2}. Similarly, the change of generators $(P_\rho,x^\lambda)\shortrightarrow(P_\rho,\hat{x}^\lambda_s)$, where
  \begin{gather}  \label{Atlh2}
\hat x^{i}_{s}=x^{i}e^{\left( 1-s\right)hP_0} , \qquad \hat x^{0}_{s} =
x^{0}- hsx^{k}P_{k}
\end{gather}%
illustrates Proposition~\ref{prop3} and gives rise to Heisenberg representation acting in the vector space $\mathfrak{X}^n[[h]]$ as well as its Hilbert space extension acting in ${\cal L}^2(\mathbb{R}^n, dx^n)[[h]]$ provided that $P_k=-\imath\partial_k$.

\subsection*{Jordanian family of twists providing $\boldsymbol{\kappa}$-Minkowski spacetime}

Jordanian twists have the following form \cite{BP}:
\begin{gather*}  
 \mathfrak{J}_{r }=\exp \left(J_{r}\otimes \sigma_r \right),
\end{gather*}
where $J_{r}=\imath(\frac{1}{r}D-L^0_0)$ with a numerical factor $r\neq 0$ labeling dif\/ferent twists and $\sigma_r =\ln (1-h rP_0)$.
Jordanian twists are related with
Borel subalgebras $\mathfrak{b}^{2}(r)=\{J_r,P_{0}\}\subset \mathfrak{igl}(n,\mathbb{R})$
which, as a matter of fact, are isomorphic to the 2-dimensional solvable Lie algebra $\mathfrak{an}^1$: $[J_r, P_{0}]=P_0$.
Direct calculations show that, regardless of the value of $r$, twisted
commutation relations (\ref{tcr}) take the form of that for $\kappa$-Minkowski spacetime
(\ref{kM1}).
The corresponding classical $r$-matrices are the following\footnote{Now, for dif\/ferent values of the parameter $r$ classical $r$-matrices are not the same.}:
\begin{gather}\label{jt2}
\mathfrak{r}_J=\mathfrak{J}_{r}\wedge P_0={1\over r}D\wedge P_0-L^0_0\wedge P_0.
\end{gather}
Since in the generic case we are dealing with $\mathfrak{igl}(n)$-twists (see~\cite{BP} for exceptions), we shall write deformed coproducts and antipodes in terms of generators $\{L_{\nu }^{\mu },P_{\mu }\}$. The deformed coproducts read as follows~\cite{BP}:
\begin{gather*}
\Delta _{r}\left( P_{0}\right) =1\otimes P_{0}+P_{0}\otimes e^{\sigma _{r}},\qquad
 \Delta _{r}\left( P_{k}\right) =1\otimes P_{k}+P_{k}\otimes e^{-\frac{1}{r}\sigma _{r}},
\\
\Delta _{r}\left( L_{k}^{m}\right) =1\otimes L_{k}^{m}+L_{k}^{m}\otimes 1,\qquad
\Delta _{r}\big( L_{0}^{k}\big) =1\otimes
L_{0}^{k}+L_{0}^{k}\otimes e^{\frac{r+1}{r}\sigma _{r}},
\\
 \Delta _{r}\left( L_{k}^{0}\right) =1\otimes L_{k}^{0}+L_{k}^{0}\otimes e^{-%
\frac{r+1}{r}\sigma _{r}}-\imath hrJ_{r}\otimes P_{k}e^{-\sigma _{r}},\\
\Delta _{r}\left( L_{0}^{0}\right) =1\otimes L_{0}^{0}+L_{0}^{0}\otimes 1
-\imath hrJ_{r}\otimes P_{0}e^{-\sigma _{r}},
 \end{gather*}
where
\begin{gather*}
e^{\beta\sigma_r}=\left(1-arP_0\right)^\beta=\sum_{m=0}^\infty \frac{%
h^m}{m!}\beta^{\underline m}(- rP_0)^m
\end{gather*}
and $\beta^{\underline m}=\beta(\beta-1)\cdots (\beta-m+1)$ denotes the
so-called falling factorial. The antipodes are:
\begin{gather*}
S_r\left( P_{0}\right) =-P_{0}e^{-\sigma_r }, \qquad S_r\left( P_{k }\right)
=-P_{k }e^{\frac{1}{r}\sigma_r },
\\
S_r\big( L_{0}^{k}\big) =-L_{0}^{k}e^{-\frac{r+1}{r}\sigma_r}, \qquad
S_r\left( L_{k}^{0}\right) =-\left(L_{k}^{0}+\imath h rJ_rP_k\right)e^{\frac{r+1}{r}\sigma_r },
\\
S_r\left( L_{0}^{0}\right) =-L_{0}^{0}-\imath hrJ_rP_0, \qquad S_r\left(
L_{k}^{m}\right)=-L_{k}^{m}.
\end{gather*}
Now using twisted coproducts and classical action
one can obtain by smash product construction (for f\/ixed value of the parameter~$r$) of extended
 position-momentum-$\mathfrak{gl}(n)$ algebra with the following crossed commutators:
\begin{gather}\label{WeylJord}
\left[ \hat{x}^{\mu },P_{0}\right]_r =\imath \delta^\mu
_0e^{\sigma _{r}}=\imath \delta^\mu_0(1-hrP_0)
, \qquad\left[ \hat{x}^{\mu },P_{k}\right]_r
=\imath\delta^\mu_k(1-hrP_0)^{-\frac{1}{r}},
\\
\nonumber
\left[ \hat{x}^{\mu },L_{k}^{m}\right]_r =\imath\hat{x}^{m}\delta^\mu_k,\qquad
\big[ \hat{x}^{\mu },L_{0}^{k}\big]_r =\imath\hat{x}
^{k}\delta^\mu_0(1-hrP_0)^{\frac{r+1}{r}},
\\
\nonumber
\left[ \hat{x}^{\mu },L_{k}^{0}\right]_r =\imath\hat{x}
^{0}\delta^\mu_k(1-hrP_0)^{-\frac{r+1}{r}}+\imath h\big(\hat{x}^{p}\delta_p^\mu
-r\hat{x}^{0}\delta^\mu_0\big)P_{k}(1-hrP_0)^{-1},
\\
\nonumber
\left[\hat{x}^{\mu },L_{0}^{0}\right]_r =\imath \hat{x}^{0}\delta^\mu
_0-\imath h\big( -\hat{x}^{k}\delta_k^\mu+r\hat{x}^{0}\delta^\mu
_0\big) P_{0}(1-hrP_0)^{-1}
\end{gather}
supplemented by $\mathfrak{igl}(n)$ (\ref{igl}) and $\kappa$-Minkowski  relations~(\ref{kMhat}).
One can see that relations (\ref{kMhat}), (\ref{WeylJord}) generate $r$-deformed phase space $\mathfrak{W}^n[[h]]^\mathcal{F}$.
Heisenberg realization is now in the following form:
\begin{gather}  \label{Jtlh}
  \hat{x}^{i}_{r}=x^{i}\left( 1-raP_0\right)^{-\frac{1}{r }}  \qquad
\mbox{and}\qquad   \hat{x}^{0}_{r}=x^{0}(1-raP_0).
\end{gather}
It is interesting to notice that above formulas~(\ref{Jtlh}) take the same form before and after Heisenberg realization. Moreover one can notice that commutation relation~(\ref{WeylJord}) can be reached by nonlinear change (\ref{Jtlh}) of generators: $(P_\rho,x^\lambda)\shortrightarrow(P_\rho,\hat{x}^\lambda_r)$. This illustrates Propositions~\ref{prop2},~\ref{prop3} for the Jordanian case. Again in the Heisenberg realization the classical $r$-matrices~(\ref{jt2}) coincide with Poisson bivector~(\ref{kM2}). Only for the case $r=-1$ the covariance group can be reduced to one-generator (dilatation) extension of the Poincar\'e algebra~\cite{Ballesteros,BP}.

\section[Different models of $\kappa$-Minkowski spacetime as a covariant $\kappa$-Poincar\'e  quantum space and corresponding DSR algebras]{Dif\/ferent models of $\boldsymbol{\kappa}$-Minkowski spacetime as a covariant\\ $\boldsymbol{\kappa}$-Poincar\'e  quantum space and corresponding DSR algebras}\label{section4}

As it is the well-known quantum deformations of the Lie algebra are controlled by classical $r$-matrices satisfying classical Yang--Baxter (YB) equation: homogeneous or inhomogeneous.
In the case of $r$-matrices satisfying homogeneous YB equations the co-algebraic sector is
twist-deformed while algebraic one remains classical~\cite{Drinfeld}.
Additionally, one can also apply existing twist tensors to relate Hopf module-algebras in order to obtain quantized, e.g., spacetimes (see \cite{BP,MSSKG,Meljanac0702215} as discussed in our previous section for quantizing Minkowski spacetime). For inhomogeneous $r$-matrices one applies Drinfeld--Jimbo (the so-called standard) quantization instead.

\begin{remark}\label{DJ}
Drinfeld--Jimbo quantization algorithm relies on
simultaneous deformations of the algebraic and coalgebraic sectors and
applies to semisimple Lie algebras \cite{Drinfeld,Jimbo}. In particular, it
implies existence of classical basis for Drinfeld--Jimbo quantized algebras.
Strictly speaking, the Drinfeld--Jimbo procedure cannot be applied to the
Poincar\'{e} non-semisimple algebra which has been obtained by the
contraction procedure from the Drinfeld--Jimbo deformation of the
anti-de~Sitter (simple) Lie algebra $\mathfrak{so}(3,2)$. Nevertheless,
quantum $\kappa$-Poincar\'{e} group shares many properties of the original
Drinfeld--Jimbo quantization. These include existence of classical basis, the
square of antipode and solution to specialization problem~\cite{BP2}. There is no cocycle twist related with the Drinfeld--Jimbo deformation. The Drinfeld--Jimbo quantization has many non-isomorphic forms (see e.g.~\cite{Klimyk,Chiari}).
\end{remark}

\begin{remark}
Drinfeld--Jimbo quantization can be considered from the more general point of view, in the so-called quasi bialgebras framework. In this case more general cochain twist instead of ordinary cocycle twist can be used together with a coassociator in order to perform quantization. Cochain twists may lead to non-associative star multiplications~\cite{non-ass}.
\end{remark}
In the classical case the physical symmetry group of Minkowski spacetime is Poincar\'{e} group and in deformed case analogously quantum $\kappa$-Poincar\'{e}  group should be desired symmetry group for quantum $\kappa$-Minkowski spacetime \cite{Z,MR}. However $\kappa$-Minkowski module algebra studied in the previous section has been obtained as covariant space over the $\U_{\mathfrak{igl}}^\mathcal{F}[[h]]$ Hopf algebra. Moreover, presented twist constructions
do not apply to Poincar\'{e} subalgebra. This is due to the fact that $\kappa$-Poincar\'{e} algebra is a quantum deformation of Drinfeld--Jimbo type corresponding to inhomogeneous classical $r$-matrix
\[
 \mathfrak{r}=  N_i\wedge P^i
 \] satisfying modif\/ied Yang--Baxter equation
\[
[[\mathfrak{r}, \mathfrak{r}]]=M_{\mu\nu}\wedge P^\mu\wedge P^\nu.
\]
As it was noticed before, one does not expect to obtain $\kappa$-Poincar\'{e} coproduct by cocycle twist. Therefore $\kappa$-Minkowski spacetime  has to be introduced as a covariant quantum space in a way without using twist and twisted star product from the previous section.

\subsection[$\kappa$-Minkowski spacetime and $\kappa$-Poincar\'{e} Hopf algebra with ``$h$-adic'' topology]{$\boldsymbol{\kappa}$-Minkowski spacetime and $\boldsymbol{\kappa}$-Poincar\'{e} Hopf algebra\\ with ``$\boldsymbol{h}$-adic'' topology}\label{section4.1}

 In this section we f\/irst recall  $\kappa$-Poincar\'{e} Hopf algebra as determine in its classical basis\footnote{Possibility of def\/ining $\kappa$-Poincar\'{e} algebra in a classical basis has been under debate for a long time, see, e.g.,  \cite{Kos, KosLuk, GNbazy, Group21, Meljanac0702215, BP2}.}. Next using ``$h$-adic'' universal $\kappa$-Minkowski spacetime algebra we  obtain ``$h$-adic'' DSR algebra via smash product construction  with the classical action.

We take the Poincar\'{e} Lie algebra $\mathfrak{io}(1,3)$ provided with a convenient choice of ``physical'' generators $(M_i,N_i,P_\mu)$\footnote{From now one we shall work with physical dimensions $n= 4$, however generalization to arbitrary dimension~$n$ is obvious.}
\begin{gather}  \label{L1}
[ M_{i},M_{j}] = \imath \epsilon _{ijk}M_{k},\qquad [ M_{i},N_{j}] =
\imath \epsilon _{ijk} N_{k},\qquad [N_{i},N_{j}] = -\imath \epsilon _{ijk} M_{k},\\
[P_\mu,P_\nu]=\lbrack M_{j},P_{0}]=0,\qquad [M_{j},P_{k}]=\imath \epsilon _{jkl}P_{l},\qquad
[N_{j},P_{k}] =- \imath \delta _{jk}P_{0},\nonumber\\
[N_{j},P_{0}]=-\imath P_{j}.  \label{L2}
\end{gather}
The structure of the Hopf algebra can be def\/ined on $\U_{\mathfrak{io}(1,3)}[[h]]$
by establishing deformed coproducts of the generators~\cite{BP2}
\begin{gather*}
\Delta _{\kappa }\left( M_{i}\right) =\Delta _{0}\left( M_{i}\right)=M_i\otimes 1+1\otimes M_i,\\
\Delta _{\kappa }\left( N_{i}\right) =N_{i}\otimes
1+\left(h P_{0}+\sqrt{1-h^{2}
P^{2}}\right)^{-1}\!\otimes N_{i}-h\epsilon _{ijm}
P_{j}\left(h P_{0}+\sqrt{1-h^{2}
P^{2}}\right)^{-1}\!\otimes M_{m},\\
\Delta _{\kappa }\left( P_{i}\right) =P_{i}\otimes
\left(hP_{0}+\sqrt{1-h^{2}
P^{2}}\right)+1\otimes P_{i},  \\
\Delta _{\kappa }\left( P_{0}\right) =P_{0}\otimes
\left(h P_{0}+\sqrt{1-h^{2}
P^{2}}\right)+\left(hP_{0}+\sqrt{1-h^{2}
P^{2}}\right)^{-1}\otimes P_{0}\\
\hphantom{\Delta _{\kappa }\left( P_{0}\right) =}{}
+hP _{m}\left(hP_{0}+\sqrt{1-h^{2}
P^{2}}\right)^{-1}\otimes P^{m},
\end{gather*}
and the antipodes
\begin{gather*}
S_{\kappa }(M_{i})=-M_{i},\qquad S_\kappa(N
_{i})=-\left(hP_{0}+\sqrt{1-h^{2}
P^{2}}\right)N_{i}-h\epsilon _{ijm}P_{j}%
M_{m},
\\
S_\kappa(P_{i})=-P_{i}\left(hP_{0}+\sqrt{1-h^{2}%
P^{2}}\right)^{-1},\qquad S_\kappa(P
_{0})=-P_{0}+h\vec{P}^{2}\left(hP_{0}+\sqrt{1-h^{2}
P^{2}}\right)^{-1},
\end{gather*}
where $P^{2}\doteq P_{\mu }P^{\mu
}\equiv\vec{P}^{2}-P_0^2$, and $\vec{P}^{2}=P
_{i}P^{i}$. One sees that above expressions are  formal
 power series in the formal parameter $h$, e.g.,
\[
 \sqrt{1-h^{2}P^{2}}=\sum_{m\geq 0}\frac{(-1)^m}{m!h^{2m}}\left(\frac{1}{2}\right)^{\underline{m}}
P^{2m}.
\]
This determines celebrated $\kappa$-Poincar\'{e} quantum group\footnote{We shall follow traditional terminology calling $\U_{\mathfrak{io}(1,3)}[[h]]$ $\kappa$-Poincar\'{e} with parameter $h$ of $[{\rm lenght}]=[{\rm mass}]^{-1}$ dimension.} in a classical basis as a Drinfeld--Jimbo deformation equipped with $h$-adic topology \cite{BP2} and from now on we shall denote it as $\U_{\mathfrak{io}(1,3)}[[h]]^{\rm DJ}$.
 This version of $\kappa$-Poincar\'{e} group is ``$h$-adic'' type and is considered as a traditional approach. The price we have to pay for it is that the deformation parameter cannot be determined, must stay formal, which means that it cannot be related with any constant of Nature, like, e.g., Planck mass or more general quantum gravity scale. In spite of not clear physical interpretation this version of $\kappa$-Poincar\'{e} Hopf algebra has been widely studied since its f\/irst discovery in~\cite{Luk1}.

\textbf{``$\boldsymbol{h}$-adic'' universal $\boldsymbol{\kappa}$-Minkowski spacetime and ``$\boldsymbol{h}$-adic'' DSR algebra.}
Since $\kappa$-Poincar\'{e} Hopf algebra presented above is not obtained by the twist deformation one needs a~new construction of $\kappa$-Minkowski spacetime as a $U_{\mathfrak{io}(1,3)}[[h]]^{\rm DJ}$ (Hopf) module algebra.

We are in position to introduce, following Remark~\ref{Poisson2} and Example~\ref{example8}, $\kappa$-Minkowski spacetime as a universal $h$-adic  algebra
$\U_h(\mathfrak{an}^3)$ with def\/ining relations\footnote{We take Lorentzian metric $\eta_{\mu\nu}={\rm diag}(-,+,+,+)$ for rising and lowering indices, e.g., $\X_\lambda=\eta_{\lambda\nu}\X^\nu$.}:
\begin{gather}\label{kMcal}
[\X_0, \X_i]=-\imath h \X_i,\qquad [\X_j, \X_k]=0
\end{gather}
Due to universal construction there is a $\mathbb{C}[[h]]$-algebra epimorphism of $\U_h(\mathfrak{an}^3)$ onto $\mathfrak{X}^4[[h]]^\mathcal{F}$ for any $\kappa$-Minkowski twist~$\mathcal{F}$.

Before applying smash product construction one has to assure that  $\U_h(\mathfrak{an}^3)$ is $\U_{\mathfrak{io}(1,3)}[[h]]^{\rm DJ}$ ($\kappa$-Poincar\'{e}) covariant algebra. It can be easily done by exploiting  the classical action
(see Example~\ref{ortho} and equation~(\ref{ClassAction}))
and by checking out consistency conditions similar to those introduced in Example~\ref{example2}, equation~(\ref{smash2}).
\begin{remark}\label{shifted} As it was already noticed that the algebra $\U_h(\mathfrak{an}^{3})$ is dif\/ferent than $\U_{\mathfrak{an}^3}[[h]]$ (see Remark \ref{Poisson2}). Assuming $P_\mu \triangleright \X^\nu=\imath a \delta^\nu_\mu$ and $[\X^0, \X^k]=\imath b \X^k$ one gets from (\ref{smash2}) and the $\kappa$-Poincar\'{e} coproduct the following relation: $b=-ah$. Particulary, our choice $a=-1$ (cf.~(\ref{ClassAction})) does imply $b=h$. In contrast $b=1$ requires $a=h^{-1}$ what is not possible for formal parameter~$h$.
This explains why the classical action cannot be extended to the unshifted generators $X^\nu$ and
entire algebra $\U_{\mathfrak{an}^3}[[h]]$. The last one seemed to be the most natural candidate for $\kappa$-Minkowski spacetime algebra in the $h$-adic case.
\end{remark}

With this in mind one can def\/ine now a DSR algebra as a crossed product extension of $\kappa$-Minkowski and $\kappa$-Poincar\'{e} algebras (\ref{L1})--(\ref{kMcal}). It is determined  by the following $\U_h(\mathfrak{an}^3)\rtimes \U_{\mathfrak{io}(1,3)}[[h]]^{\rm DJ}$ cross-commutation relations:
\begin{gather}\label{L3}
[ M_{i},\X_{0}]=0\quad \lbrack N_{i},%
\X_{0}]=-\imath \X_{i}-\imath h N_{i},
\\
\label{L4}
[ M_{i},\X_{j}]=\imath \epsilon _{ijk}\X
_{k},\qquad [ N_{i},\X_{j}]=-\imath \delta _{ij}%
\X_{0}+\imath h\epsilon _{ijk}M_{k},
\\ \label{L7}
[P_k, \X_0]=0 , \qquad [P_k, \X_j]=-\imath\delta_{jk}\left(hP_0
+\sqrt{1-h^{2}P^2}\right),
\\  \label{L8}
[P_0, \X_j]=-\imath h P_j , \qquad [P_0, \X_0]=\imath\sqrt{1-h^{2}P^2}.
\end{gather}

\begin{remark}
Relations (\ref{L3}), (\ref{L4}) can be rewritten in a covariant form:
\begin{gather*}  
 \left[ M_{\mu\nu}, \X_\lambda\right] = \imath \eta_{\mu\lambda}\X_\mu
-\imath \eta_{\nu\lambda}\X_{\mu} -\imath  h a_\mu M_{\nu\lambda}+\imath
ha_\nu M_{\mu\lambda},
\end{gather*}
where $a_\mu=\eta_{\mu\nu}a^\nu$, $(a^\nu)=(1, 0, 0, 0)$.
This form is suitable for generalization to arbitrary spacetime dimension $n$.
\end{remark}

\begin{remark}
The following change of generators:
\[
\tilde{\X}_0=\X_0,\qquad \tilde{\X}_j=\X_j+hN_j
\] provides Snyder type of commutation relation and leads to the algebra which looks like a central extension obtained in~\cite{Chryss}
(see Remark~\ref{StablForm})
\begin{gather*}\nonumber
[ P_{\mu },\tilde{\X}_{\nu }] =-i\eta_{\mu\nu }M,\qquad
[ P_{\mu },P_{\nu }] =0,\qquad
[\tilde{\X}_\mu ,\tilde{\X}_{\nu }] =\imath h^2M_{\mu \nu },
\\
[ P_{\mu },M] =0,\qquad
[\tilde{X}_\mu,M] =-\imath h^2P_{\mu },
\end{gather*}
where $M=\sqrt{1-h^2P^{2}}$ plays the role of central element and $M_{\mu\nu}$ are Lorentz generators.
\end{remark}

The center of the algebra $\U_{\mathfrak{io}(1,3)}[[h]]$ is an algebra over $\mathbb{C}[[h]]$.
Therefore one can consider a~deformation of the Poincar\'e Casimir operator $P^2$. In fact, for any power series of two variables $f(s,t)$ the element: $\mathcal{C}_f=f(P^2,h)$ belongs to the center as well. The reason of using deformed Casimir instead of the standard one is that
the standard one fails, due to noncommutativity of~$\X_\mu$, to satisfy the relation $ [P^{2},\X^{\mu }]=2P^{\mu }$. Considering deformed Casimir one has freedom to choose the form of the function~$f$. The choice
\[
\mathcal{C}_{h}^{2}=2h^{-2}\left(\sqrt{1+h^2P^2}-1\right)
\]
allows to preserve the classical properties:
 $
[M_{\mu \nu },\mathcal{C}_h^{2}]=\lbrack \mathcal{C}_h
^{2},P_{\mu }]=0$, $[\mathcal{C}_h^{2},\X_{\mu
}]=2P_{\mu }$.

The standard Poincar\'{e} Casimir  gives rise to undeformed dispersion relation:
\begin{gather}
P^{2}+m_{ph}^{2}=0,  \label{nondef}
\end{gather}
where $m_{ph}$ is mass parameter. The second Casimir operator leads to deformed dispersion relations
\begin{gather}
\mathcal{C}_h^{2}+m_{h}^{2}=0  \label{def}
\end{gather}
with the deformed mass parameter $m_h$. Relation between this two mass parameters has the following form \cite{BGMP,s1}:
\[
 m^{2}_{ph}=m_h^{2}\left(1-\frac{h^2}{4}m_{h }^{2}\right).
\]

Clearly, DSR algebra $\U_h(\mathfrak{an}^3)\rtimes \U_{\mathfrak{io}(1,3)}[[h]]^{\rm DJ}$ as introduced above is a deformation of the algebra (\ref{isog1})--(\ref{isog3}) from Example (\ref{ortho}) for the Lorentzian ($g_{\mu\nu}=\eta_{\mu\nu}$) signature; i.e., the latter can be obtained as a limit of the former when $h\rightarrow 0$. Moreover,  $\U_h(\mathfrak{an}^3)\rtimes \U_{\mathfrak{io}(1,3)}[[h]]^{\rm DJ}$ turns out to be a quasi-deformation of $\mathfrak{X}^4\rtimes \U_{\mathfrak{io}(1,3)}[[h]]$ in a sense of Propositions~\ref{prop1},~\ref{prop2}\footnote{Now we cannot make use of the twisted tensor technique from the proof of Proposition~\ref{prop1}. However, we believe, that analog of Proposition~\ref{prop1} is valid for Drinfeld--Jimbo type deformations as well.}. It means that the former algebra can be realized  by  nonlinear change of generators in the last undeformed one. To this aim we shall use explicit construction inspired by covariant Heisenberg realizations proposed f\/irstly in \cite{Meljanac0702215}:
\begin{gather}\label{naturalrealization}
\X^{\mu}=x^{\mu}\left( hp_0 +\sqrt{1-h^{2}p^2}\right)-hx^0p^\mu,\qquad M_{\mu \nu }=M_{\mu\nu},\qquad P_\mu=p_\mu
  \end{gather}%
in terms of undeformed Weyl--Poincar\'e algebra (\ref{isog1})--(\ref{isog3}) satisfying the canonical commutation relations:
\begin{gather}
\label{unWeyl}
\left[ p_{\mu },x_{\nu }\right]=-\imath \eta _{\mu \nu }, \qquad
\left[ x_{\mu },x_{\nu }\right]=
\left[ p_{\mu },p_{\nu }\right]=0.
\end{gather}
\begin{proof}The transformation $(x^\mu,M_{\mu\nu},p_\mu)\longrightarrow(\X^\mu,M_{\mu\nu},P_\mu)$ is invertible. It is subject of easy calculation that generators (\ref{naturalrealization}) satisfy DSR algebra (\ref{L3})--(\ref{L8}) provided that $(x^\mu,M_{\mu\nu},p_\mu)$ satisfy Weyl extended Poincar\'{e} algebra  (\ref{isog1})--(\ref{isog3}). This f\/inishes the proof. Note that in contrast to~\cite{Meljanac0702215} we do not require Heisenberg realization for $M_{\mu\nu}$.\footnote{We recall that in Heisenberg realization $M_{\mu\nu}=x_\mu p_\nu-x_\nu p_\mu $.} Particulary, deformed and undeformed Weyl algebras are $h$-adic isomorphic.
\end{proof}

Above statement resembles, in a sense, the celebrated Coleman--Mandula theorem~\cite{CM}: there is no room for non-trivial combination of Poincar\'{e}
and $\kappa$-Minkowski spacetime coordinates.

However there exist a huge amount of another Heisenberg realizations of the  DSR algebra $\U_h(\mathfrak{an}^3)\rtimes \U_{\mathfrak{io}(1,3)}[[h]]^{\rm DJ}$. They lead to Heisenberg representations. Particularly important for further applications is the so-called non-covariant family of realizations which is labeled by two arbitrary (analytic) functions~$\psi$,~$\gamma$. We shall write explicit form of all DSR algebra generators in terms of undeformed Weyl algebra $\mathfrak{W}^4[[h]]$-generators $(x^\mu, p_\nu)$.

Before proceeding further let us introduce a convenient notation:
for a given (analytic) function $f(t)=\sum f_n t^n$ of one variable we shall denote by
\begin{gather}
\label{h-analytic} \tilde{f}=f(-hp_0)=\sum f_n (-1)^np_0^n h^{n}\in\mathfrak{W}^4
\end{gather}
the corresponding element $\tilde f\in\mathfrak{W}^4[[h]]$.

Now generators of  deformed Weyl algebra $\U_h(\mathfrak{an}^3)\rtimes \mathfrak{T}[[h]]^{\rm DJ}$ admit the following Heisenberg realization:
\begin{gather}\label{realization}
\X^{i}=x^{i}\tilde{\Gamma}\tilde{\Psi} ^{-1} ,\qquad \X^{0}=x^{0}\tilde{\psi}  -hx^{k}p_{k}\tilde{\gamma}
\end{gather}
together with
\begin{gather}\label{P_0}
P_{i}=p_{i}\tilde{\Gamma}^{-1} , \qquad
P_{0} =h^{-1}\frac{\tilde{\Psi}^{-1}-\tilde{\Psi}}{2} +\frac{1}{2}h\vec{p}\,{}^2 \tilde{\Psi}\tilde{\Gamma}^{-2}.
\end{gather}
The momenta (\ref{P_0}) are also called Dirac derivatives \cite{Meljanac0702215,MSSKG,BP,DJMTWW}. Here
\begin{gather*}
\Psi(t)=\exp \left(
\int_{0}^{t}\frac{dt'}{\psi (t')}\right) ,\qquad \Gamma(t) =\exp \left(
\int_{0}^{t}\frac{\gamma (t') dt'}{\psi (t')}\right)
\end{gather*}
for an arbitrary choice of $\psi$, $\gamma$ such that $\psi (0)=1$.
Hermiticity of $\X^{\mu }$ requires $\psi ^{\prime }=-\frac{1}{3}\gamma $, i.e.~$\Gamma =\psi ^{-\frac{1}{3}}$.

The remaining $\U_h(\mathfrak{an}^3)\rtimes \U_{\mathfrak{io}(1,3)}[[h]]^{\rm DJ}$ generators are
just rotations and Lorentzian  boosts:
\begin{gather}\label{M_i}
M_i = \imath\epsilon_{ijk}x_jp_k=\imath\epsilon_{ijk}\X_jP_k\tilde{\Psi},
\\
N_i =  x_{i}\tilde{\Gamma}\frac{\tilde{\Psi}^{-1}-\tilde{\Psi}}{2h}-x_{0}p _{i}\tilde{\psi}\tilde{\Psi}\tilde{\Gamma}^{-1} +
\frac{\imath}{2}hx_{i}\vec{p}\,{}^2\tilde{\Psi}\tilde{\Gamma}^{-1}-hx^{k}p_{k}p_{i}
\tilde{\gamma}\tilde{\Psi}\tilde{\Gamma}^{-1}\nonumber\\
\phantom{N_i}{} =(\X_iP_0-\X_0P_i)\tilde{\Psi}.\label{N_i}
\end{gather}
The deformed Casimir operator reads as\footnote{Notice that $\mathcal{C}_h=\sum\limits_{k=0}^\infty c_kh^k$ is a well-def\/ined formal power series with entries $c_k\in\U_{\mathfrak{io}(1,3)}$.}:
\begin{gather}\label{C}
\mathcal{C}_h=h^{-2}\big(\tilde{\Psi}^{-1}+\tilde{\Psi}-2\big)-\vec{p}\,{}^2\tilde{\Psi}\tilde{\Gamma}^{-2}.
\end{gather}

\begin{remark}
It is worth to notice that besides $\kappa$-Minkowski coordinates $\X^\mu$ one can also introduce usual (commuting) Minkowski like coordinates $\tilde x^\mu\doteq \X^\mu \Psi$ which dif\/fer from $x^\mu$. The rotation and boost generators expressed above take then a familiar form:
 \begin{gather*}
  M_i = \imath\epsilon_{ijk}\tilde{x}_jP_k
\qquad\mbox{and}\qquad  N_i =(\tilde{x}_iP_0-\tilde{x}_0P_i).
\end{gather*}
\end{remark}

One can show that above realization (\ref{realization})--(\ref{N_i}) has proper classical limit: $\X^{\mu } =x^\mu$, $M_{\mu \nu }=x_\mu p_\nu-x_\nu p_\mu$, $P_\mu=p_\mu$ as $h \rightarrow 0$ in terms of the canonical momentum and position $(x^\mu, p_\nu)$ generators (\ref{unWeyl}). In contrast the variables $(\tilde x^\mu, P_\nu)$ are not canonical. Moreover, phy\-si\-cally measurable frame is provided by  the canonical variables (\ref{unWeyl}), which is important in DSR theories interpretation. For discussion of the DSR phenomenology, see, e.g.,~\cite{Liberati,ACdsrmyth} and references therein.

\begin{remark}
It is important to notice that all twisted realizations (\ref{Atlh2}), (\ref{Jtlh}) from the previous section are special case of the one above (\ref{realization}) for special choice of the functions~$\psi$,~$\gamma$. More exactly, Abelian realization (\ref{Atlh2}) one gets taking constant functions $\psi=1$ and $\gamma=s$ and Hermiticity of $\hat{x}^0$ forces $\gamma=0$.
Jordanian realization~(\ref{Jtlh}) requires $\psi=1+rt$ and $\gamma=0$ (cf.~\cite{BP} for details).
\end{remark}

\begin{example}
As an example let us choose: $\Psi=\exp(-hp_0)$, $\Gamma=\exp(-hp_0)$. Then
the representation of the Poincar\'{e} Lie algebra in this Hilbert space has the form:
\begin{gather*}
\nonumber
M_{i}=\frac{1}{2}\epsilon _{ijm}(x_{j}p _{m}-x_{m}p_{j}) ,\quad N_{i}={%
\frac{1}{2h}}x_{i} \big( e^{-2hp_{0}}-1\big) +
x_{0}p _{i}+\frac{ih}{2}x_{i}\vec{p}\,{}^2+h x^{k}p_{k} p_{i},\\
P_{i}=p_{i}e^{hp_{0}}, \qquad P_{0}=h^{-1} \sinh (hp_{0})+\frac{h}{2}\vec{p}\,{}^{2}e^{hp_{0}}.
\end{gather*}
Moreover, one can easily see that the operators $(M_i, N_i, p_\mu)$ constitute the bicrossproduct basis. Therefore, dispersion relation expressed in terms of the canonical momenta $p_\mu$ recovers the standard version of doubly special relativity theory (cf.\ formulae in \cite{AC,BACKG})
\begin{gather*}
\mathcal{C}_h=h^{-2}\big(e^{-{\frac{1}{2}}hp_{0}}-e^{{
\frac{1}{2}}hp_{0}}\big)^{2}-\vec{p}\,{}^2{e}^{hp_{0}}
\end{gather*}
implies
\begin{gather*}
m^{2}=\left[2h^{-1}
\sinh \left(\frac{hp_{0}}{2}\right)\right]^{2}-\vec{p}\,{}^{2}e^{hp_{0}}.
\end{gather*}
\end{example}

\subsection[$q$-analogs of $\kappa$-Poincar\'{e} and $\kappa$-Minkowski  algebras]{$\boldsymbol{q}$-analogs of $\boldsymbol{\kappa}$-Poincar\'{e} and $\boldsymbol{\kappa}$-Minkowski  algebras}
\label{section4.2}

 $\kappa$-Poincar\'{e} quantum group and $\kappa$-Minkowski quantum spacetime, with $h$-adic topology as described above, are not the only possible ones. One can introduce the so-called ``$q$-analog'' versions\footnote{In some physically motivated papers a phrase ``$q$-deformation'' is considered as an equivalent of Drinfeld--Jimbo deformation. In this section we shall, following general terminology of~\cite{Klimyk,Chiari}, distinguish between ``$h$-adic'' and ``$q$-analog'' Drinfeld--Jimbo deformations since they are not isomorphic.}, which allows us to f\/ix value of the parameter $\kappa$. Afterwards they become usual complex algebras without the $h$-adic topology. In the case of Drinfeld--Jimbo deformation this is always possible.
The method is based on reformulation of  a Hopf algebra in a way which allows to hide inf\/inite series. As a result one obtains a one-parameter family of isomorphic Hopf algebras enumerated by numerical parameter $\kappa$. It is deliberated as the so-called ``specialization'' procedure.
We shall describe this procedure for the case of $\kappa$-Poincar\'{e} (cf.~\cite{BP2}\footnote{Similar construction has been performed in~\cite{Stachura} in the bicrossproduct basis (see also~\cite{Z}).})
and we shall f\/ind its $\kappa$-Minkowski counterpart.
The corresponding (the so-called canonical) version of $\kappa$-Minkowski spacetime algebra seems to be very interesting from physical point of view.
In this model  $\kappa$-Minkowski algebra is a universal enveloping algebra of some solvable Lie algebra. This algebra has been also used in many recently postulated approaches to quantum gravity, e.g., Group Field Theory (see \cite{Oriti} and references therein). Not distinguishing between two versions, ``$h$-adic'' and ``$q$-analog'', has been origin of many misunderstandings in the physical literature. We shall try to clarify this point here.

The main idea behind $q$-deformation is to introduce two mutually inverse group-like elements hiding inf\/inite power series. Here we shall denote them as $\Pi_0$, $\Pi_0^{-1}$: $\Pi_0^{-1}\Pi_0=1$. Fix~\mbox{$\kappa\in \mathbb{C}^*$}. Denote by $\U_\kappa(\mathfrak{io}(1,3))$ a universal, unital and
associative algebra over complex numbers generated by eleven generators $(M_i, N_i, P_i,
\Pi_0, \Pi_0^{-1})$ with the following set of def\/ining relations:
\begin{gather}
\nonumber \Pi_0^{-1}\Pi_0=\Pi_0\Pi_0^{-1}=1,\qquad
[P _i, \Pi_0]=[M_j, \Pi_0]=0 ,\qquad [N_i, \Pi_0]=-\frac{\imath}{\kappa}P_i, 
\\
 [N_i, P_j]=-\frac{\imath}{2}\delta_{ij}\left(\kappa(\Pi_0-\Pi_0^{-1})+\frac{1}{\kappa}\vec{P}^2 \Pi_0^{-1}\right),
\label{11alg}
\end{gather}
where remaining relations between $(M_i,N_i,P_i)$ are the same as in the Poincar\'{e} Lie algeb\-ra~(\ref{L1}),~(\ref{L2}).
Commutators with $\Pi_0^{-1}$ can be easily calculated from (\ref{11alg}), e.g., $[N_i,\Pi_0^{-1}]=\frac{i}{\kappa}P_i\Pi_0^{-2}$.
Notice that all formulas contain only f\/inite powers of the numerical parameter~$\kappa$.
The  quantum algebra structure $\U_\kappa(\mathfrak{io}(1,3))$ is provided by def\/ining coproduct, antipode and counit, i.e.\ a Hopf algebra structure.
We set
\begin{gather*}
 \Delta_\kappa\left( M_{i}\right) =M_{i}\otimes 1+1\otimes M_{i},
\nonumber\\
 \Delta_\kappa\left( N_{i}\right) =N_{i}\otimes
1+\Pi_0^{-1}\otimes N_{i}-\frac{1}{\kappa}\epsilon _{ijm}
P_{j}\Pi_0^{-1}\otimes M_{m},
\\
\Delta_\kappa\left(P_{i}\right) =P_{i}\otimes
\Pi_0+1\otimes P_{i},  \qquad
\Delta_\kappa(\Pi_0)=\Pi_0\otimes \Pi_0,\qquad \Delta_\kappa
(\Pi_0^{-1})=\Pi_0^{-1}\otimes \Pi_0^{-1},\nonumber
\end{gather*}
and the antipodes
\begin{gather*}
S_\kappa(M_{i})=-M_{i},\qquad S_\kappa(N
_{i})=-\Pi_0N_{i}-\frac{1}{\kappa}\epsilon _{ijm}P_{j}M_{m},\qquad S_\kappa(P_{i})=-P_{i}\Pi_0^{-1},\\
S_\kappa(\Pi_0)=\Pi_0^{-1}, \qquad S_\kappa(\Pi_0^{-1})=\Pi_0.
\end{gather*}

To complete the def\/inition one leaves counit $\epsilon$ undeformed, i.e., $\epsilon(A)=0$ for $A\in(M_i, N_i, P _i)$ and
$\epsilon(\Pi_0)=\epsilon(\Pi_0^{-1})=1$.

For the purpose of physical interpretation, specialization of the parameter $\kappa$ makes possible its identif\/ication with some physical constant of Nature: typically it is quantum gravity scale~$M_Q$. However we do not assume a priori that this is Planck mass (for discussion see, e.g.,~\cite{BGMP}). Particularly, the value of $\kappa$ depends on a system of units we are working with. For example, one should be able to use natural (Planck) system of units, $\hbar=c=\kappa=1$, without changing mathematical properties of the underlying physical model.
From mathematical point of view, this means that the numerical value of parameter $\kappa$ is irrelevant. And this is exactly the case we are dealing with.
For dif\/ferent numerical values of $\kappa\neq 0$ the Hopf algebras $\U_\kappa(\mathfrak{io}(1,3))$ are isomorphic, i.e.\ $\U_\kappa(\mathfrak{io}(1,3))\cong\U_1(\mathfrak{io}(1,3))$. One can see that by re-scaling $P_\mu\mapsto \frac{1}{\kappa} P_\mu$.

In this version $\kappa$-Minkowski spacetime commutation relations have the Lie algebra form:
\begin{gather}\label{kM}
[X^0,X^i]=\frac{\imath}{\kappa}X^i,\qquad [X^i,X^j]=0
\end{gather}
with unshifted generators (see Remark \ref{shifted}): again the numerical factor $\kappa$ can be put to $1$ (the value of $\kappa$ is unessential) by re-scaling, i.e.\ the change of basis in the Lie algebra
$X^{0}\mapsto \kappa X^{0}$.\footnote{It was not possible in $h$-adic case with the formal parameter $h=\kappa^{-1}$.}
Relations (\ref{kM}) def\/ine fourth-dimensional solvable Lie algebra $\mathfrak{an}^3$ of rank~3, cf.\ Example~\ref{example8}.

\begin{remark}
The Lie algebra (\ref{kM}) turns out to be a Borel (i.e.\ maximal, solvable) subalgebra in de Sitter Lie algebra $\mathfrak{o}(1,4)$. It can  be seen via realization:
\[
 X^0\AC M^{04}, \qquad X^i\AC M^{0i}-M^{i4}.
 \]
This fact has been explored in \cite{Oriti,Iwasawa}.
\end{remark}
We are now in position to introduce the canonical (in terminology of~\cite{Kon}, cf.\ Remark~\ref{Poisson2}) $\kappa$-Minkowski spacetime $\M^4_\kappa=\mathcal{U}_{\mathfrak{an}^3}$ as a universal enveloping algebra of the solvable Lie algebra~$\mathfrak{an}^3$ studied before in~\cite{Oriti,Sitarz,Andrea}. In order to assure that $\M^4_\kappa$ is  $\U_\kappa(\mathfrak{io}(1,3))$-module algebra one has to check the consistency conditions (cf.\ Example~\ref{example2} and Remark~\ref{shifted}). Then utilizing the crossed product construction one obtains canonical DSR algebra $\M^4_\kappa\rtimes\U_\kappa(\mathfrak{io}(1,3))$.

\subsection*{Canonical DSR algebra}

The Hopf algebra $\U_\kappa(\mathfrak{io}(1,3))$ acts covariantly on $\M^4_\kappa$ by means of the classical action on the generators:
\begin{gather}
\label{ClassAction}
P_{i }\triangleright X^{\nu }=-\imath\delta _{i }^{\nu },\qquad
M_{\mu \nu }\triangleright X^{\rho }= \imath X_{\nu }\delta _{\mu }^{\rho }-
\imath X_{\mu }\delta _{\nu }^{\rho},\qquad
\Pi_0^{\pm 1}\triangleright X^\mu=X^\mu\mp\imath \kappa^{-1}\delta_0^\mu.
\end{gather}
As a result  cross-commutation relations determining the canonical   DSR algebra $\M^4_\kappa\rtimes\U_\kappa(\mathfrak{io}(1,3))$ take the following form:
\begin{gather}
[ M_{i},X_{0}]=0,\qquad [ N_{i},%
X_{0}]=-\imath X_{i}-\frac{\imath }{\kappa }N
_{i},  \qquad
[ M_{i},X_{j}]=\imath \epsilon _{ijl}X
_{l},\nonumber\\
 [ N_{i},X_{j}]=-\imath \delta _{ij}
X_{0}+\frac{\imath }{\kappa}\epsilon _{ijl}M_{l},\nonumber
\\
\label{phsp1}
[ P_{k},X_{0}]=0,\qquad [ P_{k},
X_{j}]=-\imath \delta _{jk}\Pi_0,\qquad
[X_i, \Pi_0]=0 ,\qquad [X_0, \Pi_0]=-\frac{\imath}{\kappa}\Pi_0.
\end{gather}
It is easy to see again that dif\/ferent values of the parameter $\kappa$ give rise to isomorphic al\-geb\-ras. One should notice that the above ``$q$-analog'' version of DSR algebra cannot be pseudo-deformation type as the one introduced in the ``$h$-adic'' case.
 A subalgebra of special interest is, of course,
  a canonical $\kappa$-Weyl algebra $\mathfrak{W}^4_\kappa$ (canonical $\kappa$-phase space) determined by the relations~(\ref{kM}) and (\ref{phsp1}).
Heisenberg like realization of the Lorentz generators in terms of $\mathfrak{W}^4_\kappa$-generators can be found as well:
\begin{gather*}
 M_{i} =\epsilon _{ijk}\left[ X^{j}\Pi _{0}^{-1}+2\kappa^{-1}X^{0}P^{j}\big( \Pi _{0}^{2}+1-\kappa^{-2}\vec{P}^{2}\big) ^{-1}\right] P_{k},
\\
N_{i}=X^{0}P^{i}\frac{3-\Pi _{0}^{-2}\big(1-\kappa^{-2}
\vec{P}^{2}\big)}{\big( \Pi _{0}+\Pi _{0}^{-1}\big(1-
\kappa^{-2}\vec{P}^{2}\big)\big) }+\frac{\kappa}{2}X^{i}\big(
1-\Pi _{0}^{-2}\big(1-\kappa^{-2}\vec{P}^{2}\big)\big).
\end{gather*}
The deformed Casimir operator reads
\begin{gather*}
\mathcal{C}_{\kappa} =\kappa^{2}\big(\Pi _{0}+\Pi
_{0}^{-1}-2\big)-\vec{P}^{2}\Pi _{0}^{-1}.
\end{gather*}
Moreover, one can notice that undeformed Weyl algebra $\mathfrak{W}^4$: $(x^\mu,p_\nu)$ is embedded in $\mathfrak{W}^4_\kappa$ via:
\begin{gather*}
  x^0=2X^0\big(\Pi_0+\Pi_0^{-1}\big(1-\kappa^{-2}\vec{P}^2\big)\big)^{-1},
\qquad
 x^i=X^i\Pi_0^{-1} +2\kappa^{-1}X^0\big(\Pi_0^2+1-\kappa^{-2}\vec{P}^2\big)^{-1}P_i,
\\
p_{0} =\frac{\kappa}{2}\big( \Pi _{0}-\Pi _{0}^{-1}\big(1-\kappa^{-2}\vec{P}^{2}\big)\big), \qquad p_i=P_i.
\end{gather*}
Finally we are in position to introduce Heisenberg representation of the canonical DSR algebra in the Hilbert space ${\cal L}^2(\mathbb{R}^4, dx^4)$. It is given by similar formulae as in the $h$-adic case (\ref{realization}), (\ref{M_i})--(\ref{C}). However here instead of $h$-adic extension $\tilde{f}$ of an analytic function $f$ (see (\ref{h-analytic})) one needs its Hilbert space operator realization
\[
\check{f}=\int f(t) dE_{(\kappa)}(t)
\]
 via spectral integral with the spectral measure $dE_\kappa (t)$ corresponding to the self-adjoint ope\-ra\-tor\footnote{Notice that $E_{(1)}=p_0$.} $E_{(\kappa)}=-\frac{\imath}{\kappa}\partial_0$, where $\kappa\in\mathbb{R}^*$.
Thus Hilbert space  representation of the canonical DSR algebra generators rewrites as (cf.~(\ref{realization}), (\ref{M_i}), (\ref{N_i})):
\begin{gather}\label{checkX}
X^{i}=x^{i}\check{\Gamma}\check{\Psi} ^{-1} ,\qquad X^{0}=x^{0}\check{\psi}  -hx^{k}p_{k}\check{\gamma},
\\
\label{checkP}
 \Pi_0=\check\Psi^{-1},\qquad
 P_{i}=p_{i}\check{\Gamma}^{-1},
\\
\label{checkM}
M_i = \imath\epsilon_{ijk}x_jp_k,
\\
\label{checkN}
N_i = \kappa x_{i}\check{\Gamma}\frac{\check{\Psi}^{-1}-\check{\Psi}}{2}-x_{0}p _{i}\check{\psi}\check{\Psi}\check{\Gamma}^{-1} +
\frac{\imath}{2\kappa}x_{i}\vec{p}\,{}^2\check{\Psi}\check{\Gamma}^{-1}
-\frac{1}{\kappa}x^{k}p_{k}p_{i}
\check{\gamma}\check{\Psi}\check{\Gamma}^{-1}
\end{gather}
with the Casimir operator:
\begin{gather}\label{checkC}
\mathcal{C}_\kappa=\kappa^2\big(\check{\Psi}^{-1}+\check{\Psi}-2\big)-\vec{p}\,{}^2\check{\Psi}
\check{\Gamma}^{-2}.
\end{gather}
Here $p_\mu=-\imath\partial_\mu$ and $x^\nu$ are self-adjoint operators acting in the Hilbert space ${\cal L}^2(\mathbb{R}^4, dx^4)$.
This leads to the St\"{u}ckelberg version of relativistic Quantum Mechanics (cf.~\cite{BP,Menski}).

\begin{remark}\label{PhaseSp}
    Alternatively, one can consider relativistic symplectic structure (cf.~(\ref{unWeyl}))
    \begin{gather}\label{RSS}
        \{x^\mu, x^\nu\}=\{p_\mu, p_\nu\}=0, \qquad \{x^\mu, p_\nu\}=\delta^\mu_\nu
    \end{gather}
 determined by the symplectic two-form $\omega=dx^\mu\wedge d p_\mu$ on the phase space $\mathbb{R}^4\times\mathbb{R}^4$. Now we can interpret formulas (\ref{M_i}), (\ref{N_i}) for the f\/ixed value $h={1\over\kappa}$ as a non-canonical transformation (change of variables) in the phase space. Thus in this new variables one gets $\kappa$-deformed phase space \cite{ncphasespace} with deformed Poisson brackets replacing the commutators in formulas~(\ref{L7}),~(\ref{L8}): $\{\;,\; \}_\kappa={1\over\imath}[\;,\; ]$. It corresponds to the so-called ``dequantization'' procedure~\cite{AF}. Conversely, the operators (\ref{checkX})--(\ref{checkC}) stand for true (Hilbert space) quantization of this deformed symplectic structure.
\end{remark}

\begin{example}
As a yet another example let us consider deformed phase space of Magueijo--Smolin model \cite{Smolin}, see also~\cite{ncphasespace}:
\begin{gather*}
 \{X^\mu, X^\nu\}=\frac{1}{\kappa}(a^\mu X^\nu-a^\nu X^\mu),\qquad
\{P_\mu, P_\nu\}=0, \qquad \{X^\mu, P_\nu\}=\delta^\mu_\nu+\frac{1}{\kappa}a^\mu P_\nu.
\end{gather*}
It corresponds to  the following change  of variable in the phase space~(\ref{RSS}):
\[
X^\mu=x^\mu-\frac{a^\mu}{\kappa}x^\nu p_\nu, \qquad P_\mu=p_\mu.
\]
 We do not know twist realization for this algebra.
\end{example}

\section{Physical consequences of DSR algebra formalism}\label{section5}

After discussing  mathematical issues involving quantum $\kappa$-Minkowski spacetime and its $\kappa$-Poincar\'{e} symmetry let us focus on some physical ones. As already mentioned it is very important to clarify when one has a physical interpretation of the mathematical description.
$\kappa$-Minkowski algebra has the deformation parameter $\kappa$ which might be understood as Planck scale or Quantum Gravity scale if one chooses correct version of algebra to work on. Then $\kappa$ might denote scale at which quantum gravity corrections become relevant, dispersion relations become deformed and ``$\kappa$'' scale should become invariant (reference independent~-- for all observers). Assuming this we end up in DSR theory interpretation. However as recently has been noticed~\cite{ACdsrmyth} $\kappa$-Minkowski/$\kappa$-Poincar\'{e} formalism is not the only one which can be appropriate for description of DSR theories. Nevertheless it seems to be very fruitful and compatible with it, becoming even more promising providing framework for recent suggestions of experimentally testing quantum gravity theories. In this fashion  describing modif\/ied dispersion relations and time delay with respect to dif\/ferent noncommutative $\kappa$-Minkowski spacetime realizations allows us to provide certain bounds on quadratic corrections, i.e.\ on quantum gravity scale (see \cite{BGMP} for complete exposition) using data from the GRB's (gamma ray bursts). It was argued that the dispersion relation following from DSR are consistent with the dif\/ference in arrival time of photons with dif\/ferent energies.
Moreover it has been observed in~\cite{smolin1} that a proper analysis of
the GRB data using dispersion relations may require more than just the
parameter given by the quantum gravity scale~$M_Q$.
We will discuss deformed dispersion relations arising from  $\kappa$-Minkowski spacetime up to quadratic order in suppression by quantum gravity scale \cite{Albert,Fermi}.
Using results obtained in previous chapters we will consider now
deformed (\ref{L3})--(\ref{L8}) DSR algebra, with its dif\/ferent
realizations leading to dif\/ferent doubly (or deformed) special relativity models and
dif\/ferent physics encoded in deformed dispersion relations.
Let us clarify this point in more detail. Deformed dispersion relation obtained with in this formalism come from deformed Klein--Gordon equation, where the role of d'Alembert operator is played  by Casimir operator of $\kappa$ Poincar\'{e} algebra:
\begin{gather}  \label{dAeq}
\left(\mathcal{C}_{\kappa} - m_\kappa^2\right)\omega_p=0,
\end{gather}
where $\omega_p=\exp{(\imath  p_\mu x^\mu)}$ represents the plane wave with the wave vector $p=(p_\mu)$.

For photons: $m=m_{\kappa }=0$ and as a consequence dispersion relations obtained from~(\ref{nondef}) and~(\ref{def}) are identical.
One can see that in general both expressions  have
the same classical limit $\frac{1}{\kappa} \rightarrow 0$ but dif\/fer by order as
polynomials in $\frac{1}{\kappa}$.
Deformed Klein--Gordon equation (\ref{dAeq}) puts constraint on wave vector $p_\mu$ in the following form of dispersion relation:
\begin{gather*}
|\vec{p}|=-\kappa\left(1-\exp \left( -\int_{0}^{-\frac{p_0}{\kappa}}\frac{%
da}{\psi (a)}\right)\right)
\exp \left( \int_{0}^{-\frac{p_0}{\kappa}}\frac{
\gamma (a)da}{\psi (a)}\right),
\end{gather*}
which takes approximate form \cite{BGMP}:
\begin{gather}  \label{disp3}
|\vec{p}|\simeq p_0 \left(1-b_1\frac{p_0}{\kappa}+b_2\frac{p_0^2}{\kappa}\right)
\end{gather}
and leads to time delay:
\begin{gather*}
\Delta t\simeq
-\frac{l}{c}\frac{p_0}{\kappa}\left(2b_1-3b_2\frac{p_0}{\kappa}\right),
\end{gather*}
where
$l$ is a distance
from the source of high energy photons.

For noncovariant realizations recalled in previous section to calculate second order corrections
one needs the following expansion:
\begin{gather*}
\psi =1-\frac{p_0}{\kappa}\psi _{1}-\frac{p_0^{2}}{\kappa^{2}}\psi _{2}+o\left(\left(-\frac{p_0}{\kappa}\right)^{3}\right),\qquad \gamma =\gamma _{0}-\frac{p_0}{\kappa}\gamma
_{1}+o\left(\left(-\frac{p_0}{\kappa}\right)^{2}\right).
\end{gather*}
This provides general formulae for the coef\/f\/icients $b_{1}$, $b_{2}$ in (\ref{disp3}):
\begin{gather*}
b_{1}=\frac{1}{2}(2\gamma _{0}-1-\psi _{1}),\\
b_{2} =\frac{1}{6}\big(1+3\psi _{1}+2\psi _{1}^{2}-\psi _{2}+3\gamma
_{0}^{2}-3\gamma _{0}+3\gamma _{1}-6\gamma _{0}\psi _{1}\big).
\end{gather*}

\textbf{Jordanian one-parameter family of Drinfeld twists} (for details see
\cite{BP,BGMP}).
The time delay for photons is:
\begin{gather*}
\Delta t\simeq
-\frac{l}{c}\frac{p_0}{\kappa}\left(-(1+r)-(1+3r+2r^2)\frac{p_0}{2\kappa}\right)=
\frac{l}{c}\frac{p_0}{\kappa}\left(1+r+\big(1+3r+2r^2\big)\frac{p_0}{2\kappa}\right).
\end{gather*}

 $\kappa $-Minkowski spacetime from
one-parameter family of \textbf{Abelian twists} \cite{BP,Bu}:
 $\psi =1$, $\gamma =s= \gamma _{0}$, $\gamma _{1}=\psi_1=\psi_2=0$ and
\begin{gather*}
 \Delta t=-\frac{l}{c}\frac{|\vec{p}|}{\kappa}\left( 2s-1+\frac{|\vec{p}|}{2\kappa}s(s-1)\right).
\end{gather*}

\subsection*{Conclusions}

This paper presents a detailed state of the art of  $\kappa$-deformations of Minkowski spacetime, underlining how it may be obtained
by twist, then insisting on two distinct versions ($h$-adic topology and $q$-analog) as well as their possible physical interpretations.
$\kappa$-Poincar\'{e}/$\kappa$-Minkowski algebras are one of the possible formalisms for DSR theories and it has been widely studied \cite{Z,MR,Luk2,Luk1,Kos,KosLuk,LukNow,Nowicki,GNbazy,Group21,Ballesteros,s1,Bu,Meljanac0702215,MSSKG,BP,BP2,BGMP, FKGN,MS2, ncphasespace}.
One can see that in this formalism the second invariant scale in DSR appears from noncommutativity of coordinates, and has the meaning analogous to speed of light in Special Relativity.
Nevertheless together with growth of popularity of this approach in DSR theories many critical remarks have appeared  \cite{DSRkrytyka}. Some of the authors, using $\kappa$-Poincar\'{e} algebra formalism, have argued its equivalence to Special Relativity \cite{DSRkrytyka}. Also recently the problem of nonlocality in varying speed of light theories appeared, however it is still under debate \cite{DSRdebateI,DSRdebate}.
Nonetheless our aim in this paper was to focus on technical aspects of $\kappa$-deformation and $\kappa$-Minkowski space time with its quantum symmetry group, which were not always clearly worked out  in the physical literature.
We started from reminding standard def\/initions and illustrative examples on crossed (smash) product construction between Hopf algebra and its module. Since $\kappa$-Minkowski spacetime is an interesting example of Hopf module algebra we provided elaborated discussion on above mentioned construction. Therefore we have used smash product construction to obtain the so called DSR algebra, which uniquely unif\/ies $\kappa$-Minkowski spacetime coordinates with Poincar\'{e} generators. Its dif\/ferent realizations are responsible for dif\/ferent physical phenomena, since we obtain dif\/ferent physical predictions, e.g., on quantum gravity scale or time delay. From the mathematical point of view we discuss two main cases (two mathematically dif\/ferent models: $h$-adic and $q$-analog) of such construction using suitable versions of $\kappa$-Poincar\'{e} and $\kappa$-Minkowski algebras. In both of them we explicitly show the form of DSR algebra and some realization  of its generators. However we also point out that one should be aware that only a $q$-analog version of $\kappa$-Poincar\'{e}/$\kappa$-Minkowski has to be considered if one wants to discuss physics.
In our paper we also remind few facts on twisted deformation, and provide $\kappa$-Minkowski spacetime as well as smash algebra, with deformed phase space subalgebras, as a result of twists: Abelian and Jordanian. Moreover we have shown that the deformed (twisted) algebra is a pseudo-deformation of an undeformed one with above mentioned Jordanian and Abelian cases as explicit examples of this theorem (Proposition~\ref{prop1}). The statement that the DSR algebra can be obtained by a non-linear change of the generators
from the undeformed one is an important result from the physical point of view, because it
provides that one can always choose a physically measurable frame related with canonical
commutation relations. Nevertheless it is only possible after either $h$-adic extension of universal enveloping algebra in $h$-adic case or introducing additional generator ($\Pi_0$) in $q$-anolog case.
 In our approach dif\/ferent DSR algebras have dif\/ferent physical consequences due to dif\/ferent realizations of $\kappa$-Minkowski spacetime. We have also introduced some realizations of deformed phase spaces (deformed Weyl algebra): $r$-deformed and $s$-deformed in Jordanian and Abelian cases respectively, and also $h$-adic and $q$-analog versions as well. Heisenberg representation in Hilbert space is also provided in all above cases. What is important in our approach that it is always possible to choose physical frame (physically measurable momenta and position) by undeformed Weyl algebra which makes clear physical interpretation~\cite{Liberati}. This implies that various realizations of DSR algebras are written in terms of the standard
(undeformed) Weyl--Heisenberg algebra  which opens the way for quantum mechanical interpretation
DSR theories in a more similar way to (proper-time) relativistic  (Stuckelberg version) Quantum Mechanics instead (in Hilbert space representations contexts\footnote{We believe that this work can be also helpful for building up a proper operator algebra formalism.}).
But with this interpretation we can go further and ask if deformed special relativity is a quantization of doubly special relativity.
As we see Deformed Special Relativity is not a complete theory yet, with open problems such as e.g. nonlocality mentioned above~\cite{DSRdebateI}. Fortunately, it has been also shown quite recently that nonlocality problem is inapplicable to DSR framework based on $\kappa$-Poincar\'e~\cite{DSRdebate}. Because of this, it seems to be very timely and interesting to deal with Hopf algebras and noncommutative spacetimes associated with them. Nevertheless in our paper we only mention DSR interpretation as one of the possible ones for phenomenology of $\kappa$-Minkowski spacetime, which is interesting and promising itself. Hence we think it is important to clarify and investigate in detail these noncommutative spacetime examples from mathematical point of view because various important technical aspects of $\kappa$-Minkowski spacetime were not always introduced in the physical literature. However we would not like to force or defend any interpretation at this stage of development.

\pdfbookmark[1]{Appendix}{app}
\section*{Appendix}

In the paper we use the notion of $h$-adic (Hopf) algebras and $h$-adic  modules, i.e.\ (Hopf) algebras and modules dressed in $h$-addic topology. Therefore, we would like to collect basic facts concerning $h$-adic topology which is required by the concept of deformation quantization (for more details see \cite[Chapter~1.2.10]{Klimyk}  and \cite[Chapter~XVI]{Kassel}). For example, in the case of quantum enveloping algebra, $h$-adic topology provides invertibility of twisting elements and enables the quantization.

Let us start from the commutative ring of formal power series $\mathbb{C}[[h]]$: it is a (ring) extension of the f\/ield of complex numbers $\mathbb{C}$ with elements of the form:
\[
\mathbb{C}[[h]]\ni a=\sum_{n=0}^\infty a_n h^n,
\]
where $a_n$ are complex coef\/f\/icients and $h$ is undetermined. One can also see
this ring as $\mathbb{C}[[h]]=\times_{n=0}^\infty\mathbb{C}$
which elements are (inf\/inite) sequences of complex numbers $(a_0,a_1,\dots,a_n,\dots)$ with powers of $h$ just ``enumerating'' the position of the coef\/f\/icient. Thus, in fact, $\mathbb{C}[[h]]$ consists of all inf\/inite complex valued sequences, both convergent and divergent in a sense of standard topology on $\mathbb{C}$. A subring of polynomial
functions $\mathbb{C}[h]$ can be identif\/ied with the set of all f\/inite sequences. Another important subring is provided by analytic functions $\mathcal{A}(\mathbb{C})$, obviously: $\mathbb{C}\subset\mathbb{C}[h]\subset\mathcal{A}(\mathbb{C})\subset\mathbb{C}[[h]]$. Slightly dif\/ferent variant of sequence construction can be applied to obtain the real out of the rational numbers. Similarly, the ring $\mathbb{C}[[h]]$ constitute substantial extension of the f\/ield $\mathbb{C}$ and specialization of the indeterminant $h$ to take some numerical value does not make sense, strictly speaking.
The ring structure is determined by addition and multiplication laws:
\[
a+b:=\sum_{n=0}^\infty(a_n+b_n)h^n,\qquad a\cdot b:= \sum_{n=0}^\infty\left(\sum_{r+s=n}^\infty a_r b_s\right)h^n.
\]
This is why the power series notation is only a convenient tool for encoding the multiplication (the so-Cauchy multiplication).
Let us give few comments on topology with which it is equipped, the so-called ``$h$-adic'' topology.
This topology is determined ``ultra-norm'' $||\cdot||_{\rm ad}$ which is def\/ined~by:
\[
\left\|\sum_{n=0}^\infty a_n h^n \right\|_{\rm ad}=2^{-n(a)},
\]
where $n(a)$ is the smallest integer such that $a_n\neq 0$  (for $a\equiv 0$ one sets $n(a)=\infty$ and therefore $||0||_{\rm ad}=0$).
It has the following properties:
\begin{gather*} 0\leq ||a||_{\rm ad}\leq 1, \qquad
||a+b||_{\rm ad}\leq \max(||a||_{\rm ad}, ||b||_{\rm ad}), \\ ||a\cdot b||_{\rm ad}=||a||_{\rm ad}||b||_{\rm ad},
\qquad ||h^k||_{\rm ad}=2^{-k}.
\end{gather*}
It is worth to notice that the above def\/ined norm is discrete (with values in inverse powers of~2).
Important property is that the element $a\in\mathbb{C}[[h]]$ is invertible if an only if $||a||_{\rm ad}=1$.\footnote{Particularly, all nonzero complex numbers are of unital ultra-norm.} Above topology makes
the formalism of formal power series self-consistent in the following sense: all formal power series becomes convergent (non-formal) in
the norm $||\cdot||_{\rm ad}$. More exactly, if $\mathbb{C}[[h]]\ni a=\sum\limits_{n=0}^\infty a_n h^n$ is a formal power series then
$||a-A_N||_{\rm ad}\rightarrow 0$, with $A_N=\sum\limits_{n=0}^{n=N} a_n h^n$ being the sequence of partial sums.
Moreover, $\mathbb{C}[[h]]$ is a topological ring, complete in $h$-adic topology;
in other words the addition and the  multiplication are continuous operations and $h$-adic Cauchy sequences are convergent to the limit which belongs to the ring.

Furthermore one can extend analogously other algebraic objects, such as vector spaces, algebras, Hopf algebras, etc. and equip them in $h$-adic topology.
Considering $V$ as a complex vector space the set $V[[h]]$ contains all formal power series $v=\sum\limits_{n=0}^\infty v_n h^n$ with coef\/f\/icients $v_n\in V$. Therefore $V[[h]]$ is a $\mathbb{C}[[h]]$-module.
More generally in  the deformation theory we are forced to work with the category of topological $\mathbb{C}[[h]]$-modules, see~\cite[Chapter~XVI]{Kassel}.
$V[[h]]$ provides an example of topologically free modules. Particularly if $V$ is f\/inite dimensional  it is also free module. Any basis $(e_1,\dots,e_N)$ in $V$ serves as a system of free generators in $V[[h]]$. More exactly
\[
\sum_{k=0}^\infty v_kh^k=\sum_{a=1}^N x^ae_a , \qquad  v_k=\sum_{a=1}^N x^a_k e_a,
\]
where the  coordinates $x^a=\sum\limits_{n=0}^\infty x^a_nh^n\in \mathbb{C}[[h]]$.
It shows that $V[[h]]$ is canonically isomorphic to~$V\otimes\mathbb{C}[[h]]$. The ultra-norm $||\cdot||_{\rm ad}$ extends to $V[[h]]$ automatically. Particulary, if $V$ is equipped with an algebra structure then the Cauchy multiplication makes $V[[h]]$ a topological algebra.

Intuitively, one can think of the quantized universal enveloping algebras
introduced in the paper as families of Hopf algebras depending
on a f\/ixed numerical parameter $h$. However this does not make sense, for an algebra def\/ined over the ring $\mathbb{C}[[h]]$. To remedy this situation, one has to introduce a new algebra, def\/ined over the f\/ield of complex numbers. However this procedure is not always possible.
One can specialize $h$ to any complex number in the case of Drinfeld--Jimbo deformation but not in the case of twist deformation. For more details and examples see e.g.~\cite[Chapter~9]{Chiari}.

\subsection*{Acknowledgements}
This paper has been supported by MNiSW Grant No. NN202 318534 . The authors  acknowledge helpful discussions with P.~Aschieri, K.~De Commer, Kumar S.~Gupta, J.~Kowalski-Glikman, J.~Lukierski, S.~Meljanac, and V.N.~Tolstoy. We would like also to thank anonymous referees for their comments improving the manuscript.

\pdfbookmark[1]{References}{ref}
\LastPageEnding

\end{document}